\documentclass{article}
\usepackage[a4paper, total={7in, 9in}]{geometry}
\usepackage[utf8]{inputenc}
\usepackage[english]{babel}
\usepackage[hidelinks]{hyperref}
\usepackage{amsmath,amssymb}
\usepackage[ruled, vlined,linesnumbered]{algorithm2e}
\usepackage{graphicx}
\usepackage{graphicx, caption, subcaption}
\usepackage{float}
\usepackage{cleveref}
\usepackage{tikz}
\usepackage{pgfplots}
\usepackage{rotating}
\usepackage{wrapfig}
\usepackage{booktabs}
\usepackage{enumitem}
\usepackage{lineno}
\usepackage{tabularx}
\usepackage{booktabs}
\usepackage{multirow}
\usepackage{siunitx}
\usepackage{shellesc}
\usepackage{stackengine}
\usepackage{authblk}

\usepackage[symbol]{footmisc}

\providecommand{\keywords}[1]
{
	\small	
	\textbf{\textit{Keywords---}} #1
}

\newcolumntype{L}[1]{>{\hsize=#1\hsize\raggedright\arraybackslash}X}%
\newcolumntype{R}[1]{>{\hsize=#1\hsize\raggedleft\arraybackslash}X}%
\newcolumntype{C}[1]{>{\hsize=#1\hsize\centering\arraybackslash}X}%

\usepackage{array}
\usepackage{arydshln}
\newlist{tabitemize}{itemize}{1}
\setlist[tabitemize]{label=\textbullet,nosep,after=\strut,align=parleft,leftmargin=*,}

\pgfplotscreateplotcyclelist{my black white}{%
	solid, every mark/.append style={solid, fill=gray}, mark=*\\%
	dotted, every mark/.append style={solid, fill=gray}, mark=square*\\%
	densely dotted, every mark/.append style={solid, fill=gray}, mark=otimes*\\%
	loosely dotted, every mark/.append style={solid, fill=gray}, mark=triangle*\\%
	dashed, every mark/.append style={solid, fill=gray},mark=diamond*\\%
	loosely dashed, every mark/.append style={solid, fill=gray},mark=*\\%
	densely dashed, every mark/.append style={solid, fill=gray},mark=square*\\%
	dashdotted, every mark/.append style={solid, fill=gray},mark=otimes*\\%
	dashdotdotted, every mark/.append style={solid},mark=star\\%
	densely dashdotted,every mark/.append style={solid, fill=gray},mark=diamond*\\%
}

\pgfplotsset{
	axis label/.append style={font=\normalsize},
	tick label style={font=\normalsize},
}
\title{A High Order Flux Reconstruction Interface Tracking Method Using Preconditioned Phase Field}

\author[1]{Jabir Al-Salami}
\author[2]{Mohamed M. Kamra}
\author[2]{Changhong Hu \footnote{Corresponding author. Email: hu@riam.kyushu-u.ac.jp}}
\affil[1]{Interdisciplinary Graduate School of Engineering Sciences, Kyushu University, Fukuoka, Kasuga, 816-8580 Japan}
\affil[2]{Research Institute for Applied Mechanics, Fukuoka, Kasuga, 816-8580 Japan}

\crefformat{figure}{Fig.~#2#1#3}
\date{}
\begin{document}
	\maketitle
  \begin{@twocolumnfalse}
	\begin{abstract}
This paper presents a simple and highly accurate method for capturing sharp interfaces moving in divergence-free velocity fields using the high-order Flux Reconstruction approach on unstructured grids.
A well-known limitation of high-order methods is their susceptibility to the Gibbs phenomenon; the appearance of spurious oscillations in the vicinity of discontinuities and steep gradients makes it difficult to accurately resolve shocks or sharp interfaces.  
In order to address this issue in the context of sharp interface capturing, a novel, preconditioned and localized phase field method is developed in this work. 
The numerical accuracy of interface normal vectors is improved by utilizing a preconditioning procedure based on the level set method with localized artificial viscosity stabilization. 
The developed method was implemented in the framework of the multi-platform Flux Reconstruction open-source code PyFR \cite{Witherden2014}.
Numerical tests in 2D and 3D conducted on different mesh types showed that the preconditioning procedure significantly improves accuracy. The results demonstrate the conservativeness of the proposed method and its ability to capture highly distorted interfaces with superior accuracy when compared to conventional and high-order VOF and level set methods. The high accuracy and locality of the proposed method offer a promising route to carrying out massively-parallel, high accuracy simulations of multi-phase, incompressible phenomena.   \\ \\
	\end{abstract}
\end{@twocolumnfalse}

\keywords{High-order methods, Interface capturing, Flux Reconstruction, PyFR}

\section{Introduction}

Simulating multi-phase phenomena requires an accurate computational fluid dynamics (CFD) scheme that is capable of capturing or tracking the interface between the different phases or immiscible fluids.
Approaches to this problem broadly fall within two frameworks:Lagrangian \textit{interface tracking methods} and Eulerian \textit{interface capturing methods}. Lagrangian approaches represent the interface explicitly using either tracking markers or particles (eg. MAC methods \cite{harlow1965}), or  a continuously deforming mesh conforming to the interface (eg. Arbitrary Lagrangian-Eulerian (ALE) methods \cite{hirt1997arbitrary, baiges2017adaptive, nithiarasu2005arbitrary}). While interface tracking methods are suitable for moderately deforming free-surfaces and fluid-structure interactions, they often struggle with severely deforming and fragmenting interfaces \cite{Marchandise2006}. Furthermore, a major draw-back associated with such methods is the computational cost incurred by frequent re-meshing. In Eulerian methods, the interface is implicitly captured by means of an auxiliary field whose value corresponds to different fluids, and that is transported and deformed by the velocity variation in the domain. Among such methods, the Volume of Fluid (VOF) and level-set methods are the most successful and widely adopted,  having been proven effective in a variety of disciplines such as flooding and violent impact problems \cite{lohner2006simulation,kleefsman2005volume, kamra2018numerical}, combustion \cite{fedkiw1999ghost,pei2015large}, atomization and evaporation \cite{Luo2019,yang2016influence}.

In VOF methods, the auxiliary variable, $\alpha$, is the volume fraction. It ranges between zero and one and represents the ratio of primary to secondary fluid volume in a discrete computational cell. Therefore, the interface between two immiscible fluids is represented by a jump in $\alpha$. If the volume fraction is advected accurately and the interface thickness is kept constant in space and time, VOF schemes have been shown to demonstrate a high level of accuracy and mass conservation. In order to achieve that, however, the most accurate VOF schemes rely on complicated and computationally intensive geometric interface reconstruction operations. Another VOF approach is the tangent hyperbolic
interface capturing (THINC). THINC methods utilize a tangent hyperbolic function to reconstruct the volume fraction field. This allows for controlling and maintaining the interface thickness using a single parameter \cite{xiao2005simple}. A common drawback of VOF methods is the difficulty of obtaining accurate estimates of interface-normal vectors and interface curvature, which are especially important in surface tension dominated flows and applications that require setting special boundary conditions at the free surface.

The phase field approach to interface tracking \cite{Olsson2005,sun2007sharp,Chiu2011} is similar to VOF in that the interface is implicitly represented by the sharp transition in the auxiliary variable (the phase field in this case).  However, instead of resorting to explicit reconstruction of the interface, a diffusion and anti-diffusion term in the transport equation of the phase field maintain a diffuse interface with controllable thickness.

Level set methods utilize a signed distance function, whereby the interface is implicitly defined as the zero-level set contour. The smooth variation of the distance function across the interface allows for accurate curvature and normal vector estimates. A major caveat of this approach, however, is the mass loss due to numerical dispersion. This issue has prompted a great deal of research efforts which attempted to enhance mass conservation by a variety of approaches, such as coupling to a VOF method \cite{Lyras2020,Qian2018}, casting the level set equations in a conservative form \cite{Olsson2005,Shervani2018}, and adaptive mesh refinement (AMR) \cite{Karakus2016}, to name a few. 

Recent trends in CFD research indicate a steadily increasing interest in high-order numerical methods. The ability of such methods to produce results with more accuracy on coarser grids when compared to conventional low order methods resulted in a growing consensus among CFD practitioners that high-order methods may constitute the basis of next-generation CFD research tools \cite{Vincent2011,wang2019comparative}. In line with this vision, some efforts in multi-phase CFD research aimed at using the superior numerical properties of high-order methods to address some of the limitations of level set and VOF approaches. Qian et. al used the high-order Weighted Essentially Non-Oscillatory(WENO) approach in their coupled THINC/level set scheme to obtain high-order polynomial representations of interfaces \cite{Qian2018}. Matsushita and Aoki \cite{Matsushita2019} employed the conservative phase field method with a 5th order WENO discretization to carry out weakly compressible free-surface simulations. 
\\ 
Attempts to utilize polynomial basis high-order methods primarily focused on leveraging the high accuracy and low numerical dissipation of the discontinuous Galerkin (DG) method to improve level-set results. Jibben and Herrmann \cite{Jibben2017} developed a Runge–Kutta DG approach to re-initialization by solving banded conservative level sets on a separate, refined Cartesian grid. This approach, however, produces a smeared interface where shearing or severe deformation occurs. Zhang and Yue \cite{Zhang2019} developed an AMR-DG method where in interface cells the gradient of the level-set function is determined by a weighted local projection scheme, but mass-loss and interface fragmentation artifacts are still present. Karakus et al. \cite{Karakus2016, Karakus2018} developed a GPU-accelerated AMR re-initialization approach to level set with artificial diffusion stabilization, yet mass loss remained an issue. To solve the issue of oscillatory level-set in reinitialization, Gross et al. \cite{Grooss2006} used a combination of artificial diffusion and exponential filtering.
Most previous DG interface-capturing research efforts shied away from VOF-like sharp interface representations. This may be attributed to high-frequency oscillations that arise in the vicinity of sharp jumps due to the Gibbs phenomenon, an issue that is emblematic of polynomial-basis high-order methods \cite{gottlieb1997gibbs}. 

This paper addresses the aforementioned challenges and proposes a novel high-order method for interface capturing using the flux reconstruction (FR) approach.
The FR approach was proposed by Hyunh \cite{Huynh2007}, and it encompasses the DG and spectral difference (SD) high-order approaches. The compact spatial stencil of the FR discretization in addition to its amenability to explicit time marching makes it particularly suitable for modern high-performance-computing hardware, which is characterized by an excess of computational power relative to memory bandwidth \cite{Witherden2014,Loppi2018}.

In order to accurately capture sharp interfaces without introducing numerical oscillations, this work proposes a novel approach based on the conservative phase field method \cite{Chiu2011}. The proposed approach leverages the accuracy of the FR discretization to capture artifact-free, sharp interfaces while maintaining good mass conservation. In order to alleviate the issue of erroneous interface curvature and normal vectors, this paper proposes a level set-based, nonlinear preconditioning equation with localized artificial viscosity stabilization. This results in a simple to implement algorithm, with minimal coupling overhead and high level of accuracy. This method, henceforth referred to as the Flux Reconstruction Preconditioned Phase Field (FR-PCPF) method, has been implemented in the framework of the open-source code PyFR \cite{Witherden2014}, which is a high-performance, multi-platform flux reconstruction code.

This paper begins with a brief review of the level set and phase field methods. This is followed by a description of spatial and temporal discretization in the framework of the flux reconstruction method and other implementation details. The proposed method is validated and aspects of its performance are assessed by carrying out 2 and 3-dimensional interfacing capturing benchmarks. The results are also compared to other previous published studies using low and high-order methods. Finally, concluding remarks are given with a brief overview of future work.

\section{Governing Equations}
Phase field based interface capturing techniques are mostly derived from either the Allen-Cahn \cite{allen1979microscopic} equation or Cahn-Hillard equation \cite{cahn1958free}, which were proposed to describe the spontaneous separation in multi-component fluid mixtures. The Allen-Cahn equation is more commonly used to model multi-phase flows since it is simpler to implement and, unlike the Cahn-Hillard equation, it does not contain high-order derivatives. 
The conventional Allen-Cahn equation reads

\begin{equation}
	\frac{\partial \phi}{\partial t} = \gamma \left[ \nabla^2 \phi - \frac{F^\prime (\phi)}{\epsilon^2}  \right]
\label{allencahneq}
\end{equation}

where $\phi$ is the phase field variable in the range of zero to one, $\gamma$ and $\epsilon$ are parameters controlling the width of the interface,  and $F(\phi) = \frac{1}{2} \left( \phi^2 \left(1-\phi^2 \right)\right)$ is the so-called double-well potential. The fact that original equation does not guarantee mass conservation motivated attempts to enforce conservation through a variety of approaches; such as using Lagrange multipliers \cite{rubinstein1992nonlocal,brassel2011modified} and by formulating the equation in a conservative form \cite{jeong2017conservative}. The approach presented in this paper is based on the conservative phase field approach  by Chiu and Lin \cite{Chiu2011}, which incorporates an anti-diffusive term to Eq.\ref{allencahneq}, yielding

\begin{equation}
	\frac{\partial \phi}{\partial t} + \mathbf{u} \cdot \nabla \phi = \gamma \left[ \nabla^2 \phi - \frac{\phi(1-\phi)(1-2\phi)}{2\epsilon} - |\nabla \phi| \nabla \cdot \mathbf{n}\right].
\end{equation}

where $\mathbf{n}= \frac{\nabla \phi}{|\nabla \phi|}$ is the interface's unit normal vector. Using a kernel to define the phase field in terms of a distance function, $\psi$ such that $\phi = \frac{1}{2} \left[1+tanh\left(\frac{\psi}{2\epsilon}\right)\right]$, the first and second order derivatives can be expressed algebraically as follows

\begin{equation}
	|\nabla \phi|= \frac{\partial \phi}{\partial \psi} = \frac{\phi(1-\phi)}{\epsilon}
\end{equation}

\begin{equation}
|\nabla^2 \phi|= \frac{\partial^2 \phi}{\partial \psi^2} = \frac{\phi(1-\phi)(1-2\phi)}{\epsilon^2}
\end{equation}

by collecting the terms and assuming a divergence-free velocity field, the following equation is derived

\begin{equation}
\frac{\partial \phi}{\partial t} + \nabla \cdot(\phi \mathbf{u})   = \overline{\gamma} \left(\epsilon \nabla \cdot (\nabla\phi) -\nabla \cdot \left[ \phi (1-\phi) \mathbf{n} \right] \right)
\label{conservativephasefieldeq}
\end{equation}

where $\overline{\gamma}= \frac{\gamma}{\epsilon}$, $\gamma= |\mathbf{u}|_{max}$. In this work, however, it was found that a local choice of the multiplier yields better results, especially in areas of the flow with very small velocity magnitudes, rewriting Eq. \ref{conservativephasefieldeq} as

\begin{equation}
\frac{\partial \phi}{\partial t} + \nabla \cdot(\phi \mathbf{u})   = \max(\gamma, |\mathbf{u}|) \left(\epsilon \nabla \cdot (\nabla\phi) -\gamma \nabla \cdot \left[ \phi (1-\phi) \mathbf{n} \right] \right).
\label{altconservativephasefieldeq}
\end{equation}
Numerical experiments showed that the optimal parameters are $\gamma = 0.09$ and $\epsilon = 0.06 h/p$  where $h$ is the average grid size and $p$ is the order of the nodal basis used in the FR discretization, which will be discussed in the following sections.

A major issue with Eq. \ref{conservativephasefieldeq} is the ill-conditioning of the normal vector, especially away from the interface, which causes erroneous estimate of interface location and thickness. This issue is common with the conservative level-set method \cite{Shervani2018}, due to the similarity of this equation with the re-initialization equation used to maintain the hyperbolic tangent profile of the interface. In the conservative level-set literature, attempts to improve the accuracy of normal computations were made by recycling the signed distance function using inverse hyperbolic tangent \cite{zhao2014improved}, restoring the signed distance function by a fast marching method (FMM) \cite{desjardins2008accurate}, and algebraic stabilization terms \cite{Shervani2018,shukla2010interface}. Those approaches, however, do not completely eliminate spurious and oscillatory interface behavior. 

Inspired by the notion of non-linear pre-conditioning \cite{glasner2001nonlinear}, Chiu\cite{Chiu2019} replaced the right hand side of Eq. \ref{conservativephasefieldeq} with an equivalent expression in terms of the distance function in addition to a space-time Lagrange multiplier. To provide $\psi$, this approach required an additional phase field equation. \\
In this paper, we propose an alternative preconditioning method based on traditional level-set re-initialization which minimizes interface displacement and does away with the need to compute expensive Lagrangian multipliers. The normal vector in Eq. \ref{altconservativephasefieldeq} is instead computed using $\psi$, such that

\begin{equation}
	\mathbf{n}= \frac{\nabla \psi}{|\nabla \psi|}
\end{equation}

and $\psi$ is simply advected according to

\begin{equation}
	\frac{\partial \psi}{\partial t} + \nabla \cdot(\psi \mathbf{u}) = 0.
\end{equation}

In the beginning of the simulation, and every \textit{N} time-steps, $\psi$ is reset using

\begin{equation}
	\psi_0 = \phi-\frac{1}{2}
\end{equation} 

to form the initial condition of the following stabilized preconditioning equation 

\begin{equation}
	\frac{\partial \psi}{\partial \tau} + sgn(\psi_0)(1-|\nabla \psi|) = \nabla\cdot(\nu_s \nabla \psi).
	\label{lsequation}
\end{equation}

where $\tau$ is pseudo-time and is irrelevant to the physical problem, $sgn(\psi) = \tanh( \frac{\psi_0}{|\nabla \psi|})$ is a smeared sign function and $\nu_s$ is the stabilization viscosity. Eq. \ref{lsequation} is similar in form to the level-set re-initialization equation, however, it is not iterated until convergence to the signed-distance field. Instead we found it sufficient to only carry out the preconditioning procedure every $500 - 2000$ physical time-steps and only up to a prescribed norm. 

\section{Numerical Method}

\subsection{Flux Reconstruction Procedure of The Phase Field Equation}

We start by rewriting the phase field time-evolution (Eq. \ref{altconservativephasefieldeq}) and preconditioning (Eq. \ref{lsequation}) equations in the following form

\begin{equation}
	\frac{\partial \mathbf{u}}{\partial t} + \nabla \cdot \mathbf{f}(\mathbf{u},\mathbf{q})=0
\end{equation}

\begin{equation}
\mathbf{q} - \nabla \mathbf{u} =0
\end{equation}
where $\mathbf{u}= [\phi, \psi]^T$. Borrowing an analogy to the Navier-Stokes discretization, the fluxes of the phase field and preconditioning variables are split into "inviscid" and "viscous" parts $\mathbf{f}=\mathbf{f}^{(inv)}-\mathbf{f}^{(vis)}$, such that 
\begin{equation}
	\mathbf{f}^{(inv)} = \begin{bmatrix}
	\phi v_x & \phi v_y & \phi v_z\\
	\psi v_x & \psi v_y & \psi v_z
	\end{bmatrix},
\end{equation} 
\begin{equation}
\mathbf{f}^{(vis)} = \max(\gamma, |\mathbf{v}|) \begin{bmatrix}
\epsilon \frac{\partial \phi}{\partial x} - \gamma \left[ \frac{\phi (1-\phi)}{|\nabla \psi|} \frac{\partial \psi}{\partial x}\right] & \epsilon \frac{\partial \phi}{\partial y} - \gamma \left[ \frac{\phi (1-\phi)}{|\nabla \psi|} \frac{\partial \psi}{\partial y}\right] & \epsilon \frac{\partial \phi}{\partial z} - \gamma \left[ \frac{\phi (1-\phi)}{|\nabla \psi|} \frac{\partial \psi}{\partial z}\right]\\
0 & 0 & 0
\end{bmatrix}.
\end{equation}

Let the physical domain be divided into $|\Omega|$ elements, where $\Omega$ is the set of non-overlapping elements, such that

\begin{equation}
	\Omega = \bigcup_{n=0}^{|\Omega|-1} \Omega_n \text{ and } \bigcap_{n=0}^{|\Omega|-1} \Omega_n= \emptyset
\end{equation}
where for the sake of brevity we assume that $\Omega_n$ can be of any geometric shape in the dimension of the problem, such as quadrilateral or triangle in 2D, and prism, hexahedron, etc. in 3D.

The partial differential equations are satisfied inside each element, for instance

\begin{equation}
\frac{\partial u_n}{\partial t} + \nabla \cdot \mathbf{f}(u_n,\mathbf{q}_n)=0
\label{maineq},
\end{equation}

\begin{equation}
	\mathbf{q}_n - \nabla u_n=0
\end{equation}

where from this point on, we use the non-bold typface $u$ to refer to either $\phi$ or $\psi$.
In order to simplify implementation and improve computational efficiency, flux reconstruction steps are carried out in transformed element space. This is achieved by means of mapping functions that transform different element types into their respective standard elements. These standard elements are defined in transformed coordinates, $\widetilde{\mathbf{x}}$, such that

\begin{equation}
	\widetilde{\mathbf{x}} = \mathcal{M}^{-1}(\mathbf{x})
\end{equation}
\begin{equation}
\mathbf{x} = \mathcal{M}(\widetilde{\mathbf{x}})
\end{equation}

The Jacobians related to this mapping are

\begin{equation}
	\mathbf{J}_n = \frac{\partial\mathcal{M}_{ni}}{\partial \widetilde{x}_j}, \; \; \;   \mathcal{J}_n = \det(\mathbf{J}_n)
\end{equation}

\begin{equation}
\mathbf{J}^{-1}_n = \frac{\partial\mathcal{M}^{-1}_{ni}}{\partial \widetilde{x}_j}, \; \; \;   \mathcal{J}_n^{-1} = \det(\mathbf{J}^{-1}_n) = 1/\mathcal{J}_n.
\end{equation}

Using the transformation above, the flux and auxiliary variable in transformed space are 

\begin{equation}
	\widetilde{\mathbf{f}}_n(\widetilde{\mathbf{x}},t) = \mathcal{J}_n(\widetilde{\mathbf{x}})\mathbf{J}^{-1}_n(\mathcal{M}_n(\widetilde{\mathbf{x}})) \mathbf{f}_n(\mathcal{M}_n(\widetilde{\mathbf{x}}), t)
	\label{maintranseq}
\end{equation}

\begin{equation}
	\widetilde{\mathbf{q}}_n(\widetilde{\mathbf{x}},t) = \mathbf{J}^{T}_n(\widetilde{\mathbf{x}})\nabla u_n (\mathcal{M}_n(\widetilde{\mathbf{x}}), t).
	\label{maintranseq2}
\end{equation}

using $\widetilde{\nabla} = \frac{\partial}{\partial \widetilde{x}_i}$, the physical solution's time-derivative in Eq. \ref{maineq} can be expressed in terms of the transformed divergence of the flux in transformed, standard element space

\begin{equation}
	\frac{\partial u_n}{\partial t} + \mathcal{J}_n^{-1} \widetilde{\nabla} \cdot \widetilde{\mathbf{f}}_n(\widetilde{\mathbf{x}},t).
\end{equation}

In the flux reconstruction method, a set of solution points,$ \; \widetilde{\mathbf{x}}_{ni}^{(u)}$, where $1<i<N_{n}^{(u)}$, are placed within standard elements using an appropriate distribution. In this work, points are distributed according to the Gauss-Legendre points in quadrilateral elements and Wiallms-Shunn points in triangles.  Next, a nodal basis set $\mathfrak{l}_{ni}(\widetilde{\mathbf{x}})$ is constructed using $\Psi(\widetilde{\mathbf{x}})_{nj}$, which is a basis set that spans a polynomial space of order $p$ such that

\begin{equation}
	\mathit{l}_{ni} (\widetilde{\mathbf{x}}) = \mathcal{V}_{nij}^{-1} \Psi_{nj}(\widetilde{\mathbf{x}})
\end{equation}

where $\mathcal{V}_{nij}= \Psi_{ni}(\widetilde{\mathbf{x}}_{nj}^{(u)})$ are the elements of the Vandermonde matrix, and are required to satisfy the property $l_{ni}(\widetilde{\mathbf{x}}_{nj})= \delta_{ij}$.

In addition to solution points, a set of flux points ,$ \; \widetilde{\mathbf{x}}_{ni}^{(f)}$ where $1<i<N_{n}^{(f)}$, are defined on element boundaries. Flux points are distributed in transformed space in such a way that they share the same physical coordinates with the corresponding points of neighboring elements. 

The steps of solving equations \ref{maintranseq} and \ref{maintranseq2} are as follows:

\paragraph{Step 1:}
In order to compute gradients (Eq. \ref{maintranseq2}), the  continuous solution across element boundaries is first reconstructed. The discontinuous solution in the \textit{n}th element at the \textit{i}th flux point $u_{ni}^{(f)}$ is approximated by interpolating from the discontinuous solution at solution points $u_{nj}^{(u)}$ using the polynomial nodal basis

\begin{equation} \label{extrapolateSoln}
	u_{ni}^{(f)} = \sum_{j=1}^{N_{n}^{(f)}}u_{nj}^{(u)} \mathit{l}_{nj}(\widetilde{\mathbf{x}}_{ni}^{(f)})
\end{equation}

\paragraph{Step 2:}
By adopting the Local Discontinuous Galerkin approach, the common solution at flux points is then computed 

\begin{equation}
	\mathfrak{C}(u^{(f)}_{ni}) =\mathfrak{C}(u^{(f)}_{n^\prime i^\prime}) = (\frac{1}{2}- \beta)u^{(f)}_{ni}+ (\frac{1}{2}+ \beta) u^{(f)}_{n^\prime i^\prime}
\end{equation}

where $n^\prime i^\prime$ denote the coinciding flux point of a neighboring element, and the  parameter  $\beta = \pm \frac{1}{2}$

\paragraph{Step 3:}
The gradient of the continuous solution at solution points is computed using

\begin{equation} \label{continuousGrad}
	\widetilde{\mathbf{q}}_{ni}^{(u)} = \sum_{j=1}^{N_n^{(f)}}\left[ \widehat{\widetilde{\mathbf{n}}}_{nj} \cdot \widetilde{\nabla} \cdot \mathbf{g}_{nj}(\mathbf{\tilde{x}}_{ni}^{(u)}) \left(\mathfrak{C}(u^{(f)}_{nj}) - u^{(f)}_{nj} \right) \right] + \sum_{j=1}^{N_n^{(u)}} \left[ u^{(u)}_{nj} \widetilde{\nabla} \mathit{l}_{nj}^{(u)}(\mathbf{\tilde{x}}_{ni}^{(u)})\right]
\end{equation}

where $\widetilde{\mathbf{n}}$ is the normalized and transformed outward-pointing normal vector at the flux points, and $\mathbf{g}_{nj}(\mathbf{x}^{(u)})$ is the vector correction function associated with each flux point, which is used to ensure gradient continuity across element interfaces and is required to satisfy

\begin{equation}
\widehat{\widetilde{\mathbf{n}}}_{ni} \cdot \mathbf{g}_{nj}(\mathbf{x}_{ni}^{(f)}) = \delta_{ij}
\end{equation}

The transformed solution's gradient is mapped back to physical space using $\mathbf{q}_{ni}^{(u)} = \mathbf{J}^{-T} \widetilde{\mathbf{q}}_{ni}^{(u)}$, and then evaluated at flux points, using the same procedure in \ref{extrapolateSoln}

\begin{equation} \label{extrapolateGrad}
\mathbf{q}_{ni}^{(f)} = \sum_{j=0}^{N_{n}^{(f)}}\mathbf{q}_{nj}^{(u)} \mathit{l}_{nj}(\widetilde{\mathbf{x}}_{ni}^{(f)})
\end{equation}

\paragraph{Step 4:}

Using the transformed gradient found in the previous steps, the flux is evaluated at solution points in transformed space according to

\begin{equation}
\widetilde{\mathbf{f}}_{ni}^{(u)} =\mathcal{J}_{ni}\mathbf{J}^{-1}_{ni} \mathbf{f}(u_{ni}^{(u)}, \mathbf{q}_{ni}^{(u)}).
\end{equation}

Using this, the normal transformed flux at flux points is computed in a similar manner to Eqs. \ref{extrapolateSoln} and \ref{extrapolateGrad}

\begin{equation}\label{normalTransFlux}
\widetilde{f}_{ni}^{f\perp} = \sum_{j=0}^{N_{n}^{(f)}} \widehat{\widetilde{\mathbf{n}}} \cdot \widetilde{\mathbf{f}}_{ni}^{(u)} \mathit{l}_{nj}(\widetilde{\mathbf{x}}_{ni}^{(f)})
\end{equation}

\paragraph{Step 5:}

In order to find the continuous flux,  common normal flux at flux points is computed using

\begin{equation} \label{commonFlux}
\mathfrak{F}(f_{ni}^{f\perp}) = -\mathfrak{F}(f_{n^\prime i^\prime}^{f\perp}) = 	\mathfrak{F}^{(inv)}(f_{ni}^{(f)(inv)})-\mathfrak{F}^{(vis)}(f_{ni}^{(f)(vis)})
\end{equation}

Note that unlike the \textit{discontinuous normal flux} found in Eq. \ref{normalTransFlux}, the \textit{common normal flux} is found using quantities extrapolated from solution points from both sides of the interface, i.e. $ u_{in}^{(f)},u_{i^\prime n^\prime}^{(f)}, \mathbf{q}_{in}^{(f)}$ and $\mathbf{q} _{i^\prime n^\prime}^{(f)}$.  The inviscid part is simply the average of inviscid normal fluxes

\begin{equation}\label{inviscidFlux}
\mathfrak{F}^{(inv)}(f_{ni}^{(f)(inv)})= \widehat{\mathbf{n}}_{ni} \cdot \left[ \frac{f_{ni}^{(f)(inv)}+f_{n^\prime i^\prime}^{(f)(inv)}}{2} \right]
\end{equation}

and the viscous part is found using the LDG approach

\begin{equation}\label{viscousFlux}
\mathfrak{F}^{(vis)}(f_{ni}^{(f)(vis)})= \widehat{\mathbf{n}} \cdot \left[\left(\frac{1}{2}+\beta\right) \mathbf{f}_{ni}^{(f)} + \left(\frac{1}{2}-\beta\right) \mathbf{f}_{n^\prime i^\prime}^{(f)} \right] + \tau(u_{ni}^{(f)}-u_{n^\prime i^\prime}^{(f)}).
\end{equation}

Common normal fluxes are then scaled and transformed to standard element space to prepare for the final step.

\paragraph{Step 6:}

The divergence of the continuous flux is obtained via a procedure that is analogous to Eq. \ref{continuousGrad}

\begin{equation} \label{continuousDivFlux}
\widetilde{\nabla} \cdot \mathbf{\tilde{f}}_{ni}^{(u)} = \sum_{j=0}^{N_n^{(f)}}\left[ \widetilde{\nabla} \cdot \mathbf{\tilde{g}}_{nj}(\mathbf{x}_{ni}^{(u)}) \left(\mathfrak{F}(\tilde{f}_{nj}^{f\perp}) - \tilde{f}^{(f \perp)}_{nj} \right) \right] + \sum_{j=0}^{N_n^{(u)}} \left[ \mathbf{\tilde{f}}^{(u)}_{nj} \cdot \widetilde{\nabla} \mathit{l}_{nj}^{(u)}(\mathbf{\tilde{x}}_{ni}^{(u)})\right]
\end{equation}

the divergence of the flux at each solution point is then marched in time using a 4 stage Runge-Kutta scheme 
\begin{equation}
	\frac{\partial u^{(u)}_{ni}}{\partial t} = - \mathcal{J}^{-1}_{ni} \widetilde{\nabla} \cdot \widetilde{\mathbf{f}}_{ni}^{(u)}
\end{equation}

\subsection{Flux Reconstruciton Procedure for the Preconditioning Equation}

The system of equations used for preconditioning reads

\begin{equation} \label{peudoSystem1}
\frac{\partial \mathbf{\psi}}{\partial \tau} + \nabla \cdot \mathbf{g}(\psi,\mathbf{q}) - S(\psi ,\mathbf{q})=0
\end{equation}

\begin{equation}
\mathbf{q} - \nabla \mathbf{\psi} = 0
\end{equation}

where $\mathbf{g} =\mathbf{g}^{(vis)}= -\nu_s \nabla \psi $, and $S(\mathbf{u},\mathbf{q})= sgn(\psi_0)(1-|\mathbf{q}|)$. The steps taken to reconstruct the continuous $\psi$ field and computing its gradient are similar to those for the phase field equation  (steps 1 -3). However, step 4 is preceded by the intermediate step of computing the element-specific artificial viscosity amount needed to stabilize the equation by suppressing high-frequency oscillations. These oscillations arise due to the discontinuity of $\mathbf{q}$ and source term $\mathbf{S}(\psi ,\mathbf{q})$ in Eq. \ref{peudoSystem1}. The jump in $\psi$ across element boundaries is exacerbated with pseudo-time marching, especially in regions with very small gradients or whenever kinks arise. To avoid excessive diffusion in regions where it is not needed, we use localized artificial viscosity for stabilization, which is used in PyFR to treat discontinuities arising due to shocks in high Mach number compressible flows. A brief summary of the method by for finding $\nu_s$ is given here, but more details can be found in \cite{Persson2013}. 
The basic idea behind the shock sensor is to quantify the decay rate of orthogonal basis expansion coefficients. In smooth fields, these coefficients decay rapidly, but when the solution is under-resolved, posing as a discontinuity, the coefficients decay in relation to the strength of the discontinuity. 
The resolution indicator is defined as follows

\begin{equation}\label{resolutionIndicator}
s_e = \log_{10} \left( \frac{\left\langle u-\hat{u}, u-\hat{u} \right \rangle}{\left\langle u, u\right \rangle} \right)
\end{equation}

where $ \left \langle \cdot, \cdot \right \rangle$ is the inner product, $u$ is the solution described in modal basis of order $N$, and $\hat{u}$ is the truncated solution, projected onto the polynomial space of order $N-1$, as follows

\begin{equation}
	u = \sum_{i=1}^{N} \chi_i \Psi_{i}
\end{equation}

\begin{equation}
\hat{u} = \sum_{i=1}^{N-1} \chi_i \Psi_{i}
\end{equation}

$\chi_i$ being the modal expansion coefficients, found using the definition of the Vandermonde matrix, $\mathbf{u} = \mathcal{V}\mathbf{\chi}$.

A smoothed, element-wise stabilization viscosity is then found according to

\begin{equation}\label{nu_s}
\nu_s = \left\{
\begin{array}{lll}
0 & if & s_e < s_0 - \kappa \\
\frac{\nu_{max}}{2} \left( 1+ \sin \frac{\pi (s_e -s_0)}{2 \kappa}\right) & if & s_0-\kappa \leq s_e \leq s_0 + \kappa \\
\nu_{max} & if & s_e > s_0 + \kappa \\
\end{array} 
\right.
\end{equation} 

where we set $s_0 =-5$, $\kappa = 5\times 10^3$ and $\nu_{max} =0.15 h/p $

After completing the sixth stage of the flux reconstruction, the semi-discrete form of the pseudo-time evolution of the preconditioning equation is obtained

\begin{equation}\label{pseudotimesystem}
\frac{\partial \psi^{(u)}_{ni}}{\partial \tau} = - \mathcal{J}^{-1}_{ni} \widetilde{\nabla} \cdot \tilde{\mathbf{g}}_{ni} + S_{ni}.
\end{equation}

We found it sufficient to only carry out the above preconditioning procedure every $1000$ physical time-steps and iterating Eq. \ref{lsequation} until the $L^2$ norm satisfies

\begin{equation}
\sqrt{\frac{\sum_{n=1}^{|\Omega|}\sum_{i=1}^{N_n^{(u)}}{ \left(R^{(u)}_{ni} \right)^2 }}{N^{}(u)}} \leq \frac{1}{2}
\end{equation}

where the residual is

\begin{equation}
R_{ni}^{(u)} = \frac{\psi_n^2  - \psi_{n-1}^2}{\Delta \tau}
\end{equation}

 and $n$ is the pseudo-time iteration number, which typically requires $5-20$ iterations using the four stage Runge Kutta scheme.

\subsection{Boundary Conditions}
In order to evaluate the common solution $\mathfrak{C}(u^{(f)}_{ni})$ and common normal flux $\mathfrak{F}(f_{ni}^{f\perp})$ at the nodes of boundary faces, ghost states are used to define the solution $\mathfrak{B}(\mathbf{u}^{(f)}_{ni})$ and its gradient $\mathfrak{B}(\mathbf{q}^{(f)}_{ni})$  on virtual nodes on the other side of the interface. For all simulations shown in this paper, the following ghost states were used for the phase field system
\begin{equation}
\mathfrak{B}(\mathbf{u}^{(f)}_{ni}) = \mathbf{u}^{(f)}_{ni}
\end{equation}

\begin{equation}
\mathfrak{B}(\mathbf{q}^{(f)}_{ni}) = \mathbf{q}^{(f)}_{ni}
\end{equation}

\section{Validation Tests}

In this section, the proposed method is validated by comparing to popular 2D and 3D interface capturing  benchmarks and the results are compared to previous studies. In order to provide a quantitative error measure that takes into account the conservativeness of our approach and penalizes the presence of artifacts such as smearing and fragmentation, we use the following error measures. The $L_1$ error, is computed as follows

\begin{equation}
	E_{L1} =\int|\phi_i - \phi_e| d\Omega
\end{equation}

where $\phi_i$ and $\phi_e$ are the numerical and exact solutions respectively. Another measure that is widely used when quantifying interface error is the relative error

\begin{equation}
	E_r = \frac{\int|\phi_i - \phi_e| \, d\Omega }{\int\phi_e \, d\Omega }.
\end{equation}

To quantify mass conservation, we use the following measure

\begin{equation}\label{key}
E_m = \frac{\int \phi_i \, d\Omega - \int \phi_0 \, d\Omega}{\int\phi_e \, d\Omega }
\end{equation}

where $\phi_0$ is the phase field at the initial condition. Using the above-mentioned error measures also allows comparison to a large number of previously published interface capturing works.

The phase field at solution points is initialized according to 
\begin{equation} \label{initialCond}
\phi_i = \frac{1}{2} (1+ \tanh(\beta \psi^d_i))
\end{equation}
where $\psi^d$ is the signed distance function of the initial condition and $\beta$ is an interface sharpness parameter, which is set to $\beta = 3p/4h$ in order to produce a sharp yet smooth and non-oscillatory initial condition.

\subsection{Solid Body Rotation tests}
Zalesak's test problem \cite{zalesak1979fully} is used in this section to assess the accuracy and long-term time integration behavior of the proposed method. At the initial condition, a slotted disk with a radius $r=0.15$ centered at $(0.5, 0.75)$ in a unit square domain. The disk is initialized using Eq. \ref{initialCond} with the following distance function
\begin{equation}
\psi^d_i = r - \sqrt{(x-x_0)^2+(y-y_0)^2}.
\label{discEq}
\end{equation}
The slot is defined by $(|x-0.5|\leq 0.025) \land (y\leq 0.85)$. The slotted disc undergoes solid-body rotation due to the velocity field given by $(0.5-y, x-0.5)$ until it returns to its original position. This test was carried out on $100 \times 100$ and $200 \times 200$ uniform quadrilateral meshes.
 
\begin{figure}[htbp!]
	\centering
	\begin{subfigure}[b]{0.49\linewidth}
\includegraphics{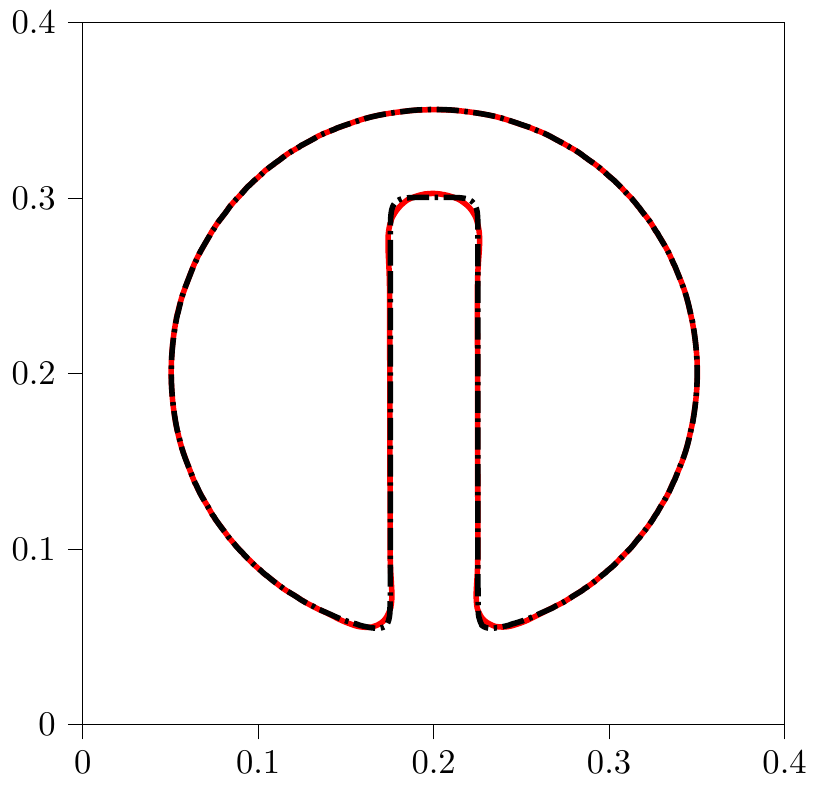}
		\caption*{Resolution : $100^2, p=2$}
	\end{subfigure}
	\begin{subfigure}[b]{0.49\linewidth}
\includegraphics{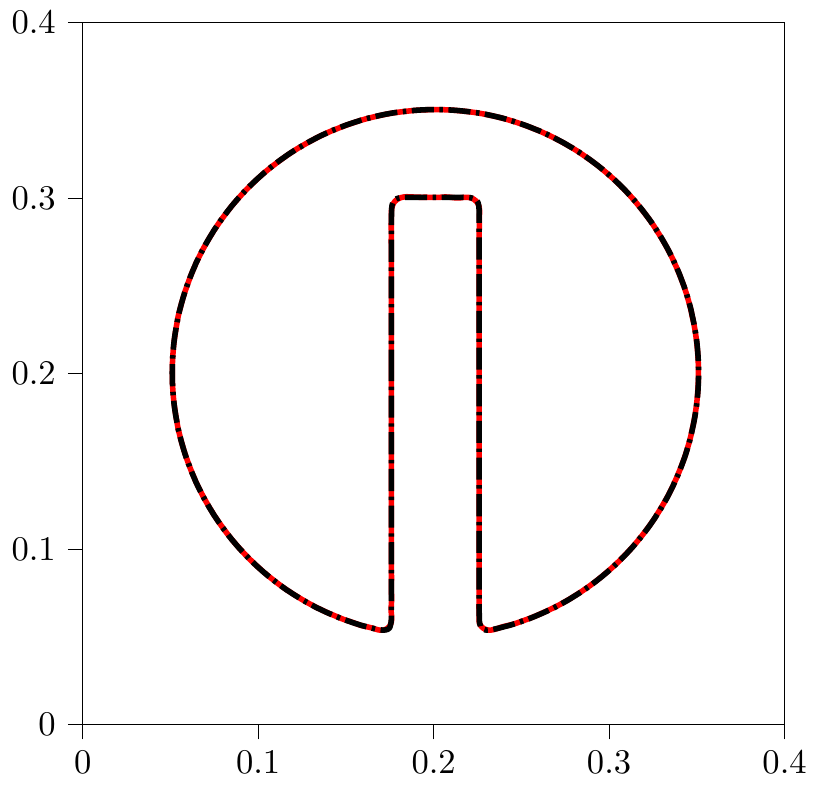}
		\caption*{Resolution : $100^2, p=4$}
	\end{subfigure}	
	\caption{Initial (black) versus final (red) $\phi=0.5$ contour of the Zalesak disc problem using two different basis orders}
	\label{zalesakDisks}
\end{figure}

 Fig. \ref{zalesakDisks} compares the initial and final shapes of Zalesak's disk on the $100^2$ mesh using $p=2$ and $p=4$. A qualitative examination of the $\phi = 0.5$ iso-line shows excellent agreement in the case of $p=2$, but some deviation from the initial condition can be seen in the corners of the slot. Increasing the basis order to $p=4$ significantly improves the ability to resolve such challenging features. Results using finer meshes or higher basis orders are not shown here as the initial and final interface contours become visually indistinguishable. 

\begin{figure}[htbp!]
	\centering
	\begin{subfigure}[b]{0.49\linewidth}
\topinset{\resizebox{0.3\linewidth}{!}{\includegraphics{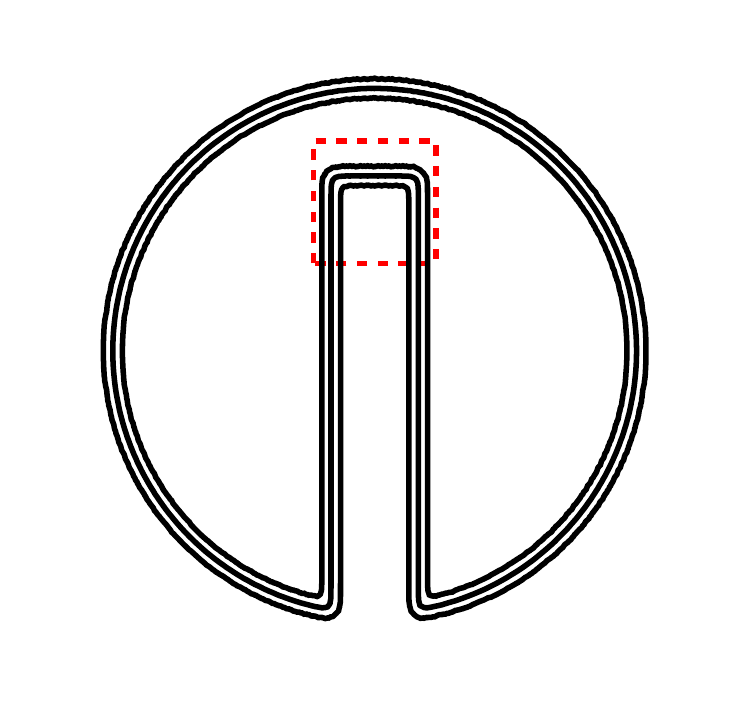}}}{\includegraphics{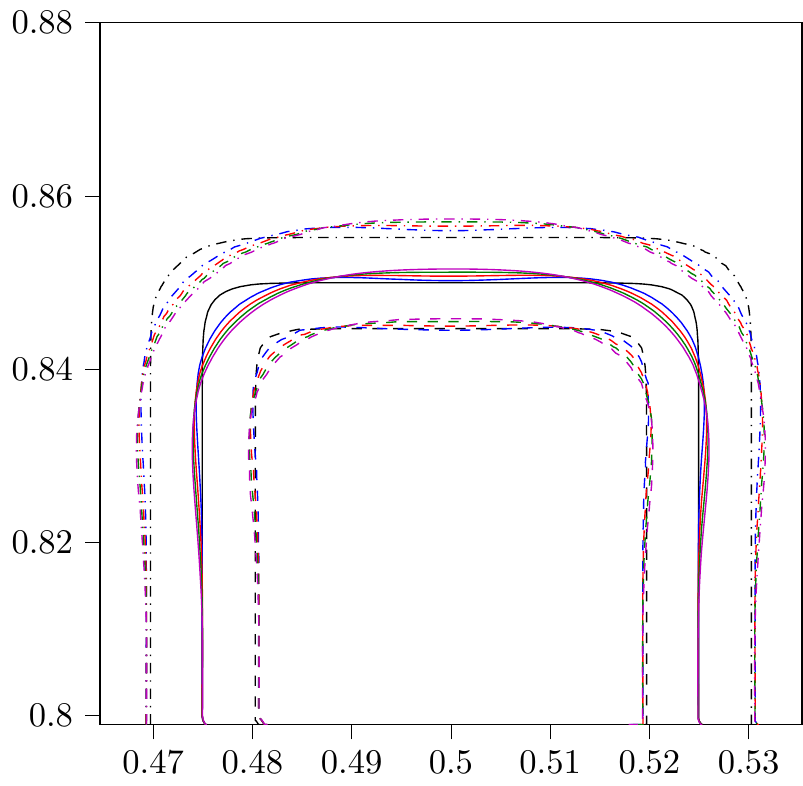}}{}{0.35\linewidth}
	\end{subfigure}
	\begin{subfigure}[b]{0.49\linewidth}
\topinset{\resizebox{0.3\linewidth}{!}{\includegraphics{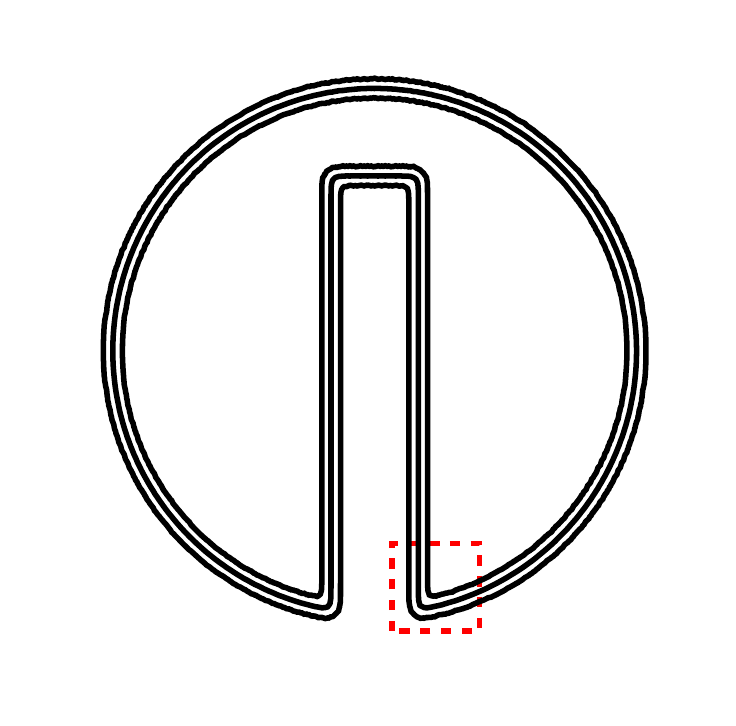}}}{\includegraphics{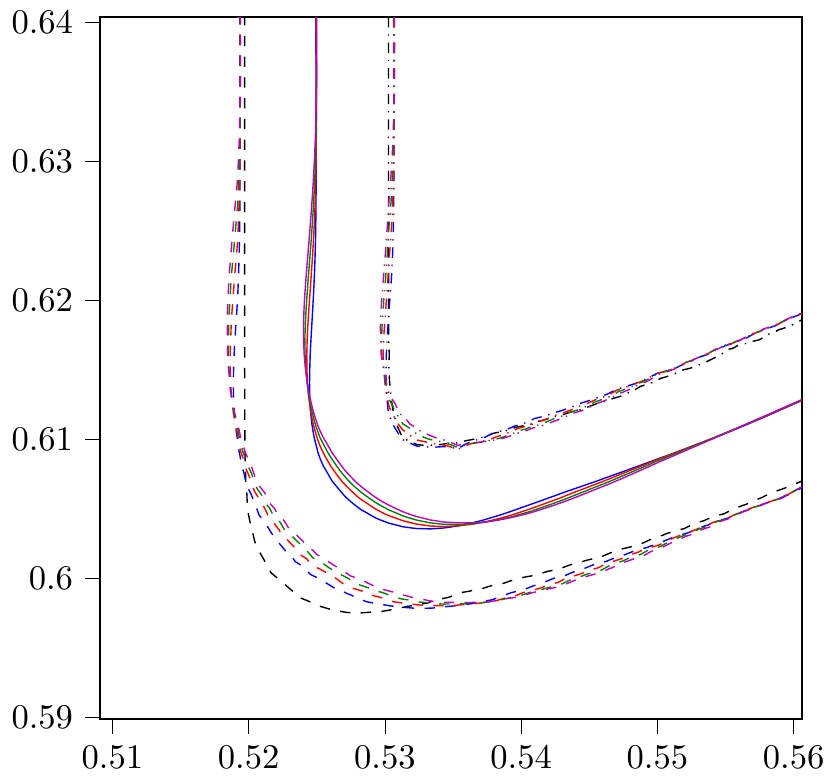}}{}{0.33\linewidth}
	\end{subfigure}
	\caption{An enlarged view of the $0.05$ (dashed), $0.5$(solid), and $0.95$ (dash-dotted) $\phi$ contours of the initial condition (black) of Zalesak's disk on the $100^2$ mesh and using $P=4$ and after 1(blue), 2(red), 3(green), 4(magenta) rotations.}
	\label{zalesakZoom}
\end{figure}

The effect of long-time integration on challenging interface features is investigated by examining the distortion in the corners of the slot after multiple rotations in \cref{zalesakZoom}.
This demonstrates that even with long-term time integration, our approach is able to preserve such challenges features and maintain a constant interface thickness.

Numerical errors using different mesh resolutions and orders are provided in Table \ref{zalesakComparisonTable} and compared to previous studies. Despite its simplicity, the FR-PCPF approach produces more accurate results when compared to more complicated approaches utilizing high order polynomial interface reconstruction. 
 
\begin{figure}
	\centering
\includegraphics{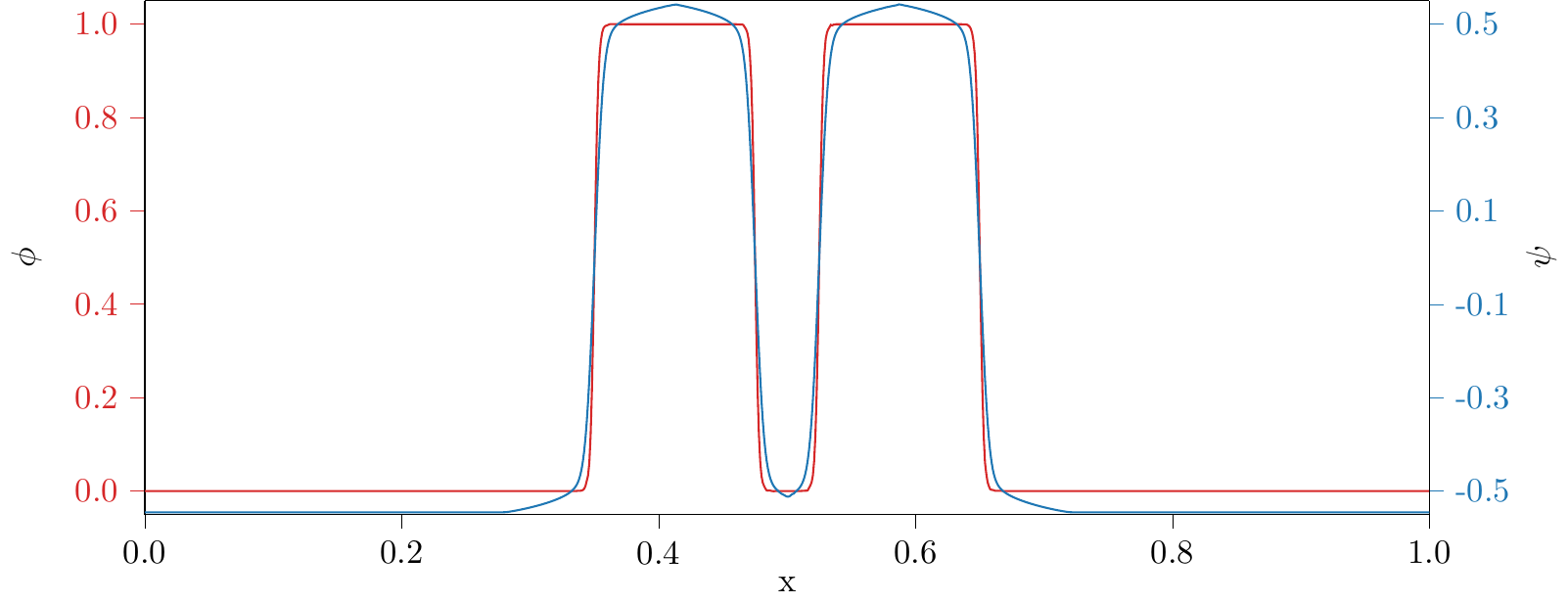}
	\caption{$\phi$ and $\psi$ plotted along the x-axis of Zalesak's disk after 1 rotation}
	\label{zalesakXline}
\end{figure}

\begin{figure}
	\centering
\includegraphics{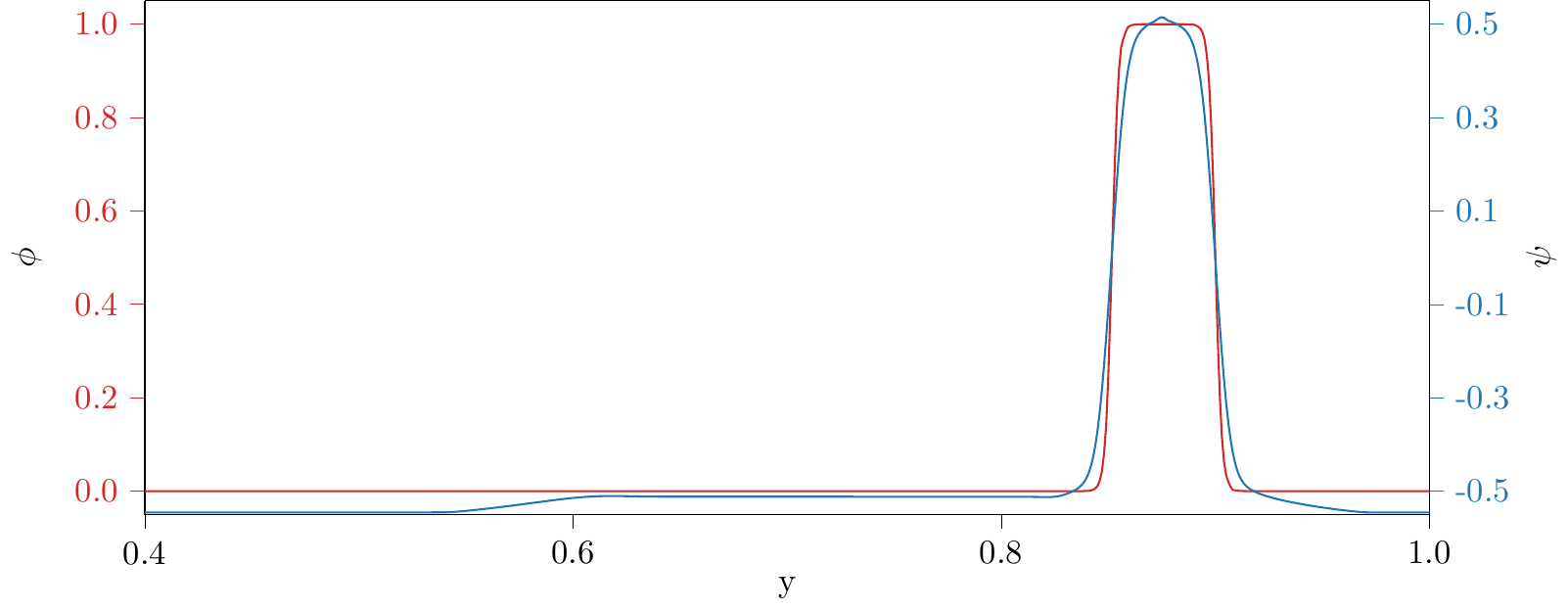}
	\caption{$\phi$ and $\psi$ plotted along the y-axis of Zalesak's disk after 1 rotation}
	\label{zalesakYline}
\end{figure}

 Figures \ref{zalesakXline} and \ref{zalesakYline} show the phase field, $\phi$ and auxiliary preconditioning variable $\psi$ plotted along the center-lines of the slotted disk at $x=0.5$ and $y=0.75$. The preconditioning procedure produces a sufficiently smooth $\psi$ field which allows for accurate normal vector computation with minimal offset between the interfaces of the phase field and auxiliary variable at $\phi = 0.5$ and  $\psi = 0$ respectively. 

We also carry out the slotted disk problem proposed by Rudman \cite{rudman1997volume}. In this problem, a slotted disk with $r=0.5$ and center at $(2.0,2.75)$ with a slot defined by $(|x-2.0|\leq 0.06) \land (y\leq 2.65)$ is placed in a $[0,4]^2$ domain with a $200 \times 200$ uniform quadrilateral mesh. The velocity field in this case is $(2-y, x-2)$. 

\begin{figure}[htbp!]
	\centering
	\begin{subfigure}[b]{0.49\linewidth}
\includegraphics{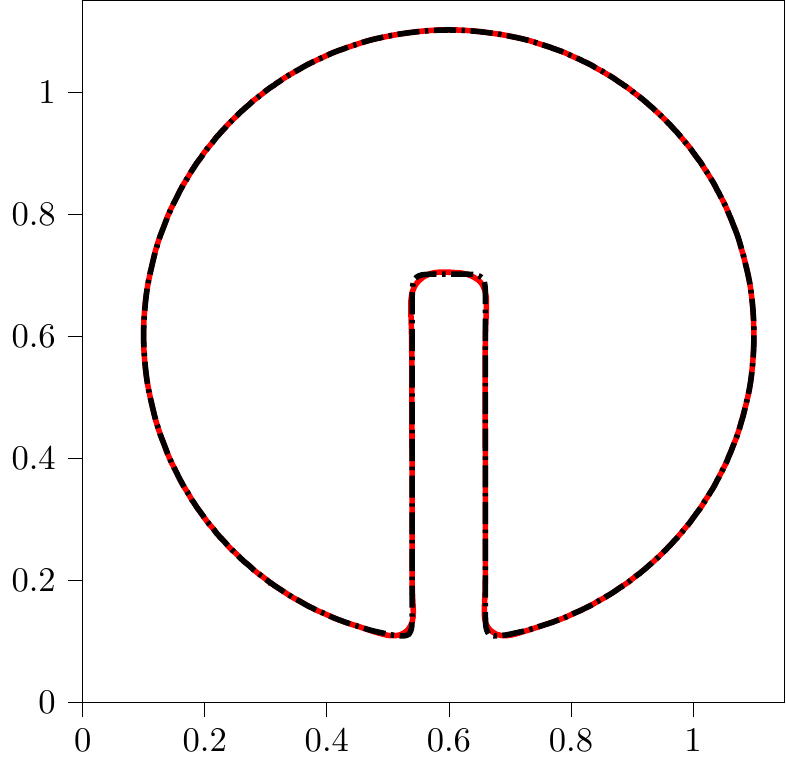}
		\caption*{Resolution : $200^2, p=2$}
	\end{subfigure}
	\begin{subfigure}[b]{0.49\linewidth}
\includegraphics{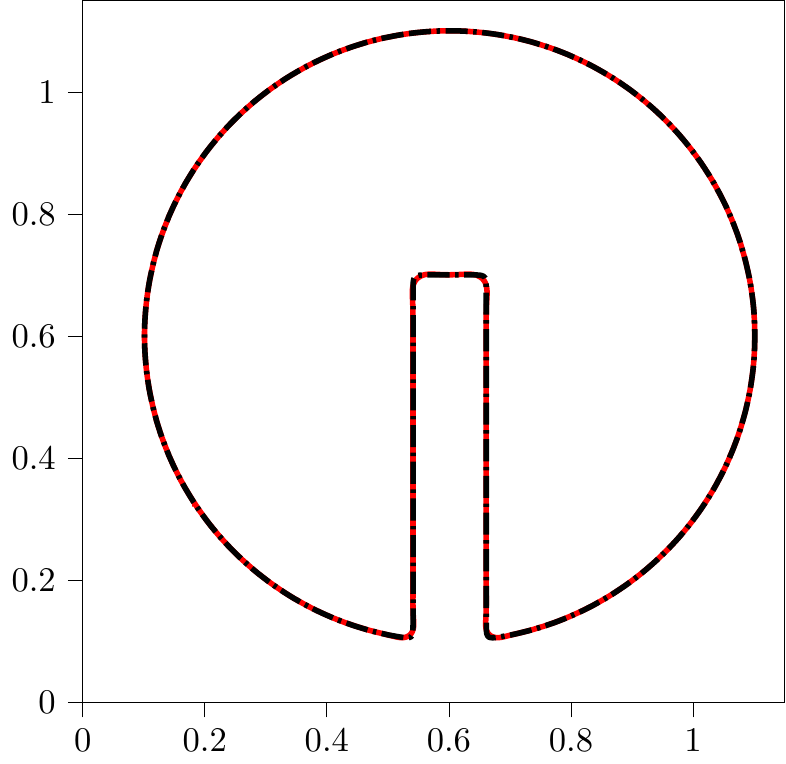}
		\caption*{Resolution : $200^2, p=4$}
	\end{subfigure}	
	\caption{Initial (black) versus final (red) $\phi=0.5$ contour of the Rudman disc problem using two different basis orders}
	\label{rudmanDisks}
\end{figure}

Fig. \ref{rudmanDisks} shows the results using $p=2$ and $p=4$, which again demonstrate the accuracy of the FR-PCPF interface capturing approach and its ability to resolve challenging interface features. This is further confirmed by quantitatively examining the error in comparison to published literature as shown in Table \ref{rudmanDikComparison}. Our results using $p=2$ outperform geometric VOF methods with a significant margin, while producing comparable results with the coupled THINC/level set approach \cite{Qian2018} method with second order polynomial interface reconstruction. Using $p\geq 3$ is sufficient to produce superior results compared to other methods.

\begin{table}[H]
	\centering
	\begin{tabularx}{0.5\textwidth}{L{0.4} L{0.25} L{0.25}}
		\toprule
		\textbf{Method}	  &    $100^2$    &     $200^2$      \\ \midrule \midrule
		MTHINC \cite{ii2012interface} &$1.61 \times 10^{-2}$ &$7.91 \times 10^{-3}$ \\
		THINC/LS(P2) \cite{Qian2018} &$1.47 \times 10^{-2}$ &$7.04 \times 10^{-3}$ \\
		THINC/LS(P4)\cite{Qian2018} &$9.98 \times 10^{-3}$ &$3.99 \times 10^{-3}$ \\ \\
		
		FR-PCPF($p=2$)  &$1.00\times 10^{-2}$ &$4.69 \times 10^{-3}$ \\
		FR-PCPF($p=3$)  &$3.49\times 10^{-3}$ &$1.54 \times 10^{-3}$\\
		FR-PCPF($p=4$)  &$2.82\times 10^{-3}$ &$9.12 \times 10^{-4}$\\
		FR-PCPF($p=5$)  &$1.63\times 10^{-3}$ &$8.08 \times 10^{-4}$\\
		\bottomrule
	\end{tabularx}
	\caption{$E_{r}$ error for the Zalesak problem on a quadrilateral mesh after one rotation }
	\label{zalesakComparisonTable}
\end{table}

\begin{table}[H]
	\centering
	\begin{tabularx}{0.5\textwidth}{L{0.75} L{0.25}}
		\toprule
		\textbf{Method}	  &  Error  \\ \midrule \midrule
		EMFPA/Youngs \cite{lopez2004volume} &$1.06 \times 10^{-2}$ \\
		EMFPA/Puckett \cite{lopez2004volume} &$9.73 \times 10^{-3}$ \\
		EMFPA/SIR \cite{lopez2004volume} &$8.74 \times 10^{-3}$ \\
		Geometric predictor-corrector \cite{cervone2009geometrical} &$9.79 \times 10^{-3}$ \\
		THINC/QQ \cite{xie2017toward} &$1.42 \times 10^{-2}$ \\
		Quadratic fit + Lagrangian\cite{scardovelli2003interface} &$4.16 \times 10^{-3}$ \\
		THINC/LS(P2) \cite{Qian2018} &$4.93 \times 10^{-3}$ \\
		THINC/LS(P4) \cite{Qian2018} &$3.77 \times 10^{-3}$ \\ \\
		FR-PCPF($p=2$)  &$4.22 \times 10^{-3}$ \\
		FR-PCPF($p=3$)  &$1.80 \times 10^{-3}$\\
		FR-PCPF($p=4$)  &$1.28 \times 10^{-3}$\\
		FR-PCPF($p=5$)  &$8.68 \times 10^{-4}$\\
		\bottomrule
	\end{tabularx}
	\caption{$E_{r}$ for the Rudman problem on a $200 \times 200$ quadrilateral mesh after one rotation }
	\label{rudmanDikComparison}
\end{table}

\subsection{Rider-Koth Vortex}
In this  benchmark case, which was first proposed in \cite{rider1998reconstructing}, a disk with a radius $r=0.15$  centered at $(0.5, 0.75)$  is initially placed within a domain of size $[0,1]^2$. The disk is then advected by the following velocity field

\begin{equation}
\begin{split}
u(x,y,t) = -\sin^2(\pi x) \sin(2\pi y) \cos (\frac{\pi t}{T}) \\
v(x,y,t) = \sin(2\pi x) \sin^2(\pi y) \cos (\frac{\pi t}{T}).
\end{split}
\end{equation}
The spatially and temporally varying velocity field described above significantly deforms the disk by stretching it into a thin spiral filament until $t=T/2$. In this second half of the test the flow is reversed, such that if a perfect scheme were to be used  the  disk would recover its initial shape at $t=T$.  In all simulations shown in this subsection the period was set to $T= 8 $.  Low-order methods are typically unable to resolve sub-grid interface features, thereby producing fragmented droplets whenever the width of the filament drops below that of the grid. However, since in the present approach  all field variables are projected to  a high-order polynomial space, sub-grid interface features can be resolved and such artifacts are not produced, provided that a sufficiently high order basis is used.

\begin{figure}[htbp!]
	\centering
	\begin{subfigure}[b]{0.45\textwidth}
\includegraphics{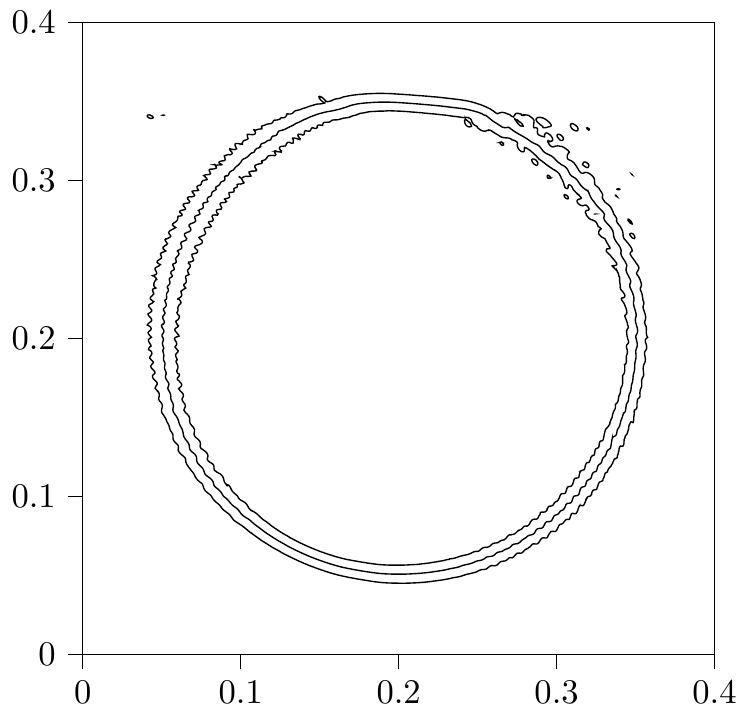}
	\end{subfigure}
	\begin{subfigure}[b]{0.45\textwidth}
\includegraphics{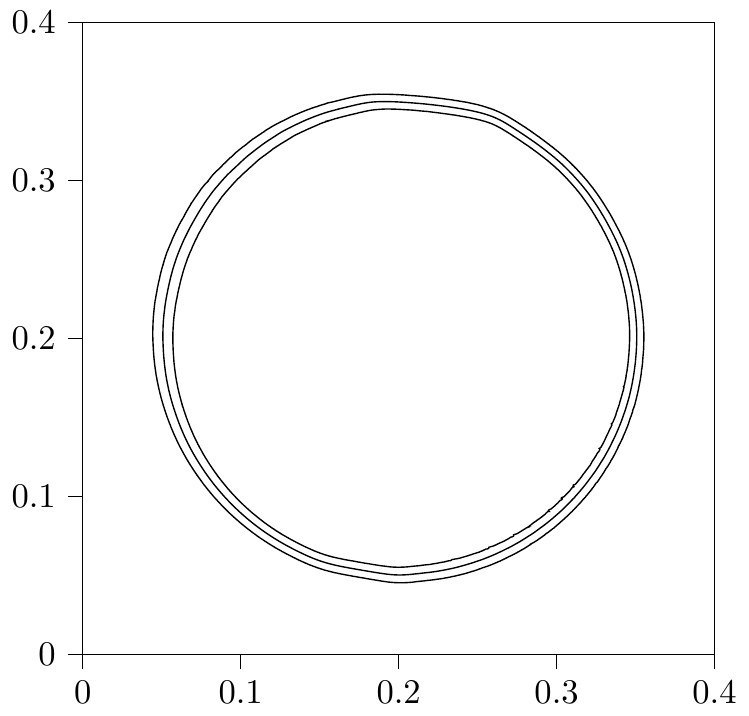}
\end{subfigure}
	\caption{Unpreconditioned (left) versus preconditioned $ \phi = 0.05,0.5,0.95 $ contour lines at $t=T$ for the Rider-Kothe vortex problem}
	\label{PCPFvsPF}
\end{figure}

Fig. \ref{PCPFvsPF}, illustrates the significant improvement in accuracy using the preconditioned phase field method, when compared to the original method. The large variations in the thickness of the interface that result from  using the original formulation are eliminated due to preconditioning. Furthermore, the proposed method also eliminates high-frequency oscillation, gaps, and fragmentation in the phase field. 

\begin{figure}[h]
  	\centering
	\begin{subfigure}[b]{0.25\textwidth}
		\includegraphics[width=\textwidth]{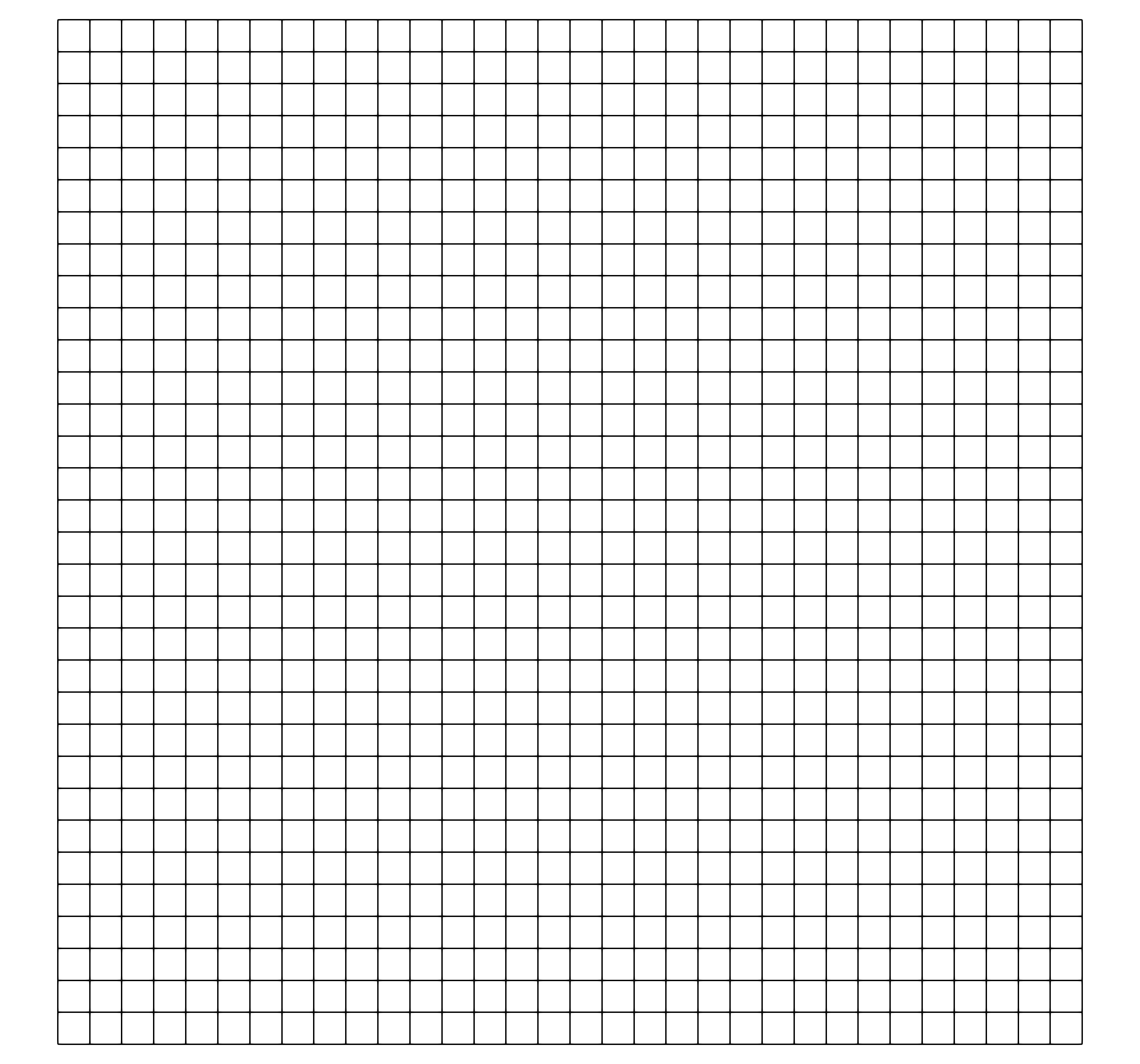}	
		\caption{}
		\label{grds_quads}
	\end{subfigure}
	\begin{subfigure}[b]{0.25\textwidth}
		\includegraphics[width=\textwidth]{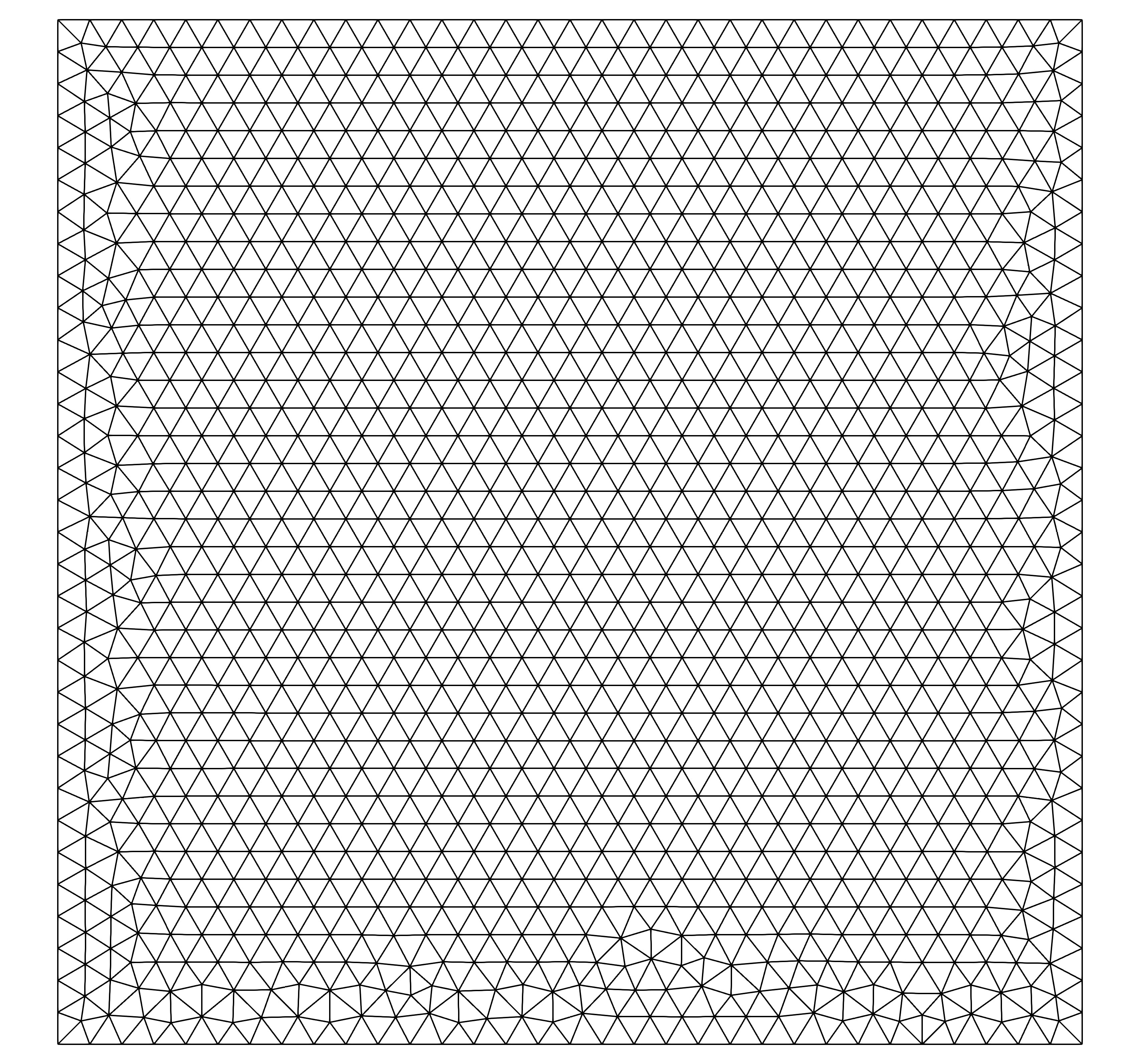}
		\caption{}
		\label{grds_trig}
	\end{subfigure}
	\begin{subfigure}[b]{0.25\textwidth}
	\includegraphics[width=\textwidth]{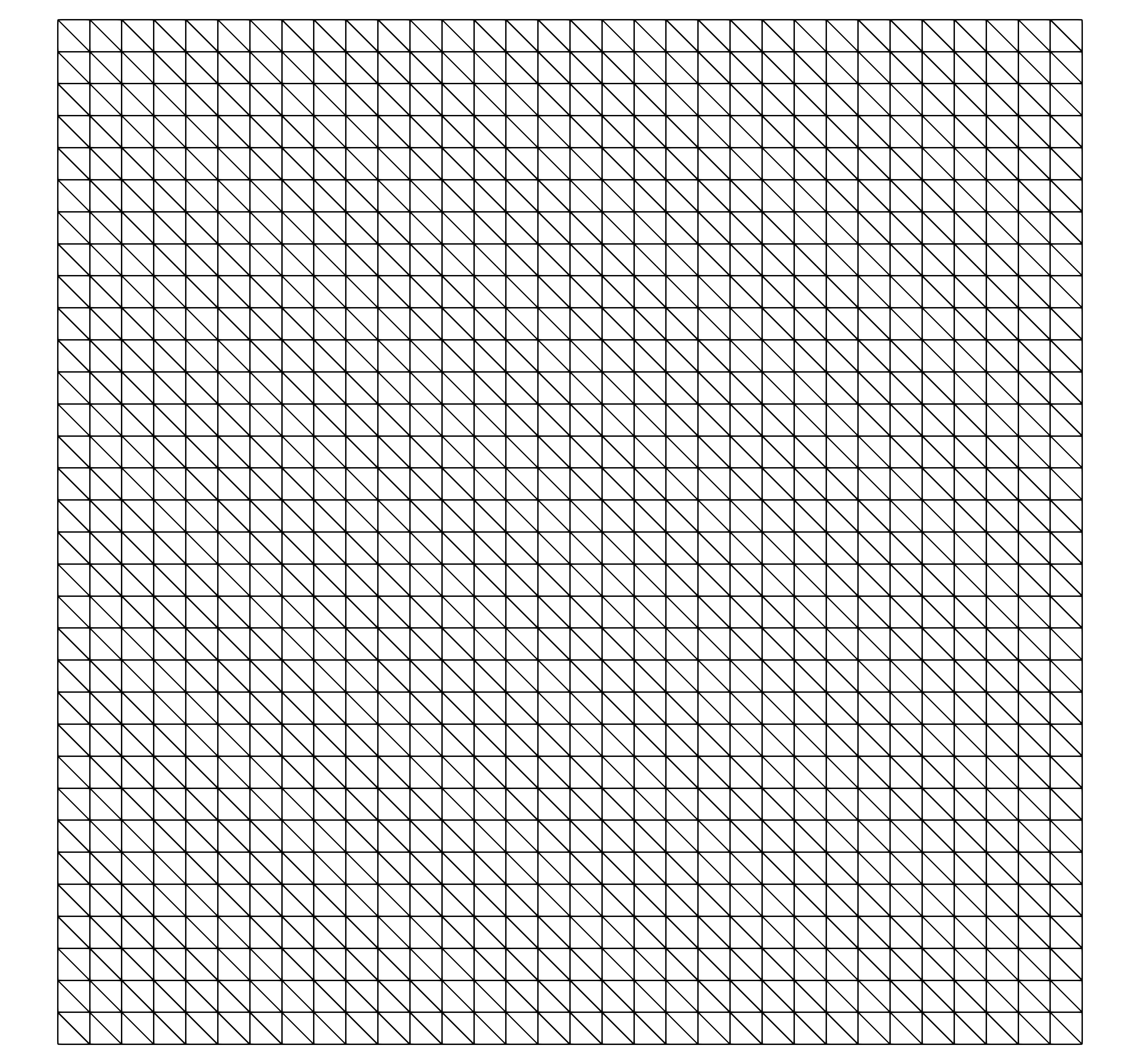}
	\caption{}
	\label{grds_trigs}
\end{subfigure}
	\caption{ Meshes used in Rider-Koth cases}
	\label{meshes}
\end{figure}

Three meshes shown in Fig. \ref{meshes}  were used for this test to demonstrate the robustness of this approach against mesh type. The three types are:
\begin{itemize}
	\item  Uniform Quadrilateral Mesh (\cref{grds_quads}): it is a uniform Cartesian mesh, which has been thoroughly investigated in literature and is used to provide a baseline for comparison with other mesh types.
	\item Nearly Equilateral Triangular Mesh (\cref{grds_trig}): it is an unstructured triangular mesh generated using the advancing-front algorithm. 

	\item Uniform Split Triangular Mesh (\cref{grds_trigs}): it is an unstructured triangular mesh produced by diagonally splitting a Cartesian mesh. 
\end{itemize}
  
Errors for different meshes and basis orders are listed in Table \ref{rvortexTable}, quantitatively demonstrating the  improvement in accuracy resulting from our proposed preconditioning method. The improvement can range from 80\% to over 300\% in some cases. The table shows that such improvement does not seem to be consistent with increasing order or base mesh resolution and could require further investigation. 

\begin{figure}[htbp!]
	\centering
	\begin{subfigure}[b]{0.45\linewidth}
\includegraphics{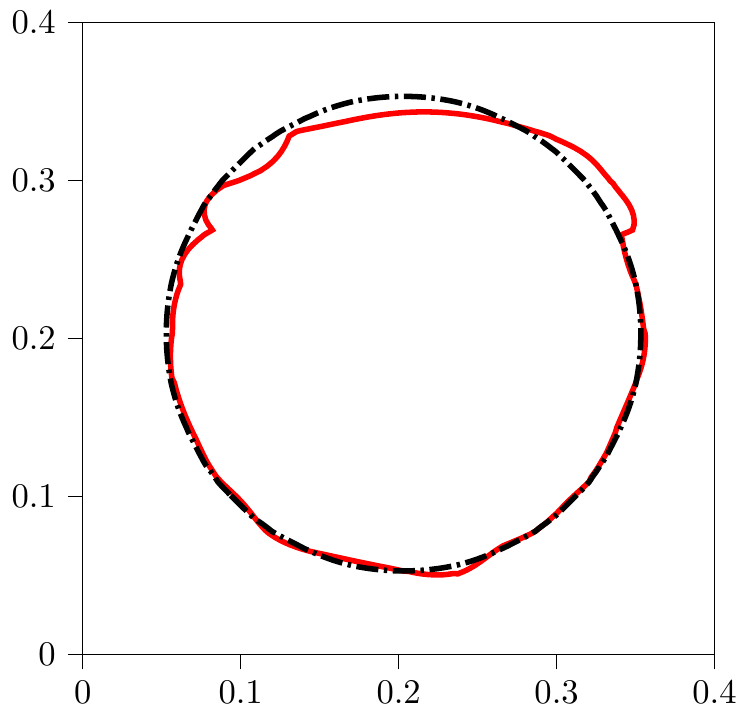}
		\caption*{Resolution : $32^2, p=2$}
	\end{subfigure}
	\begin{subfigure}[b]{0.45\linewidth}
\includegraphics{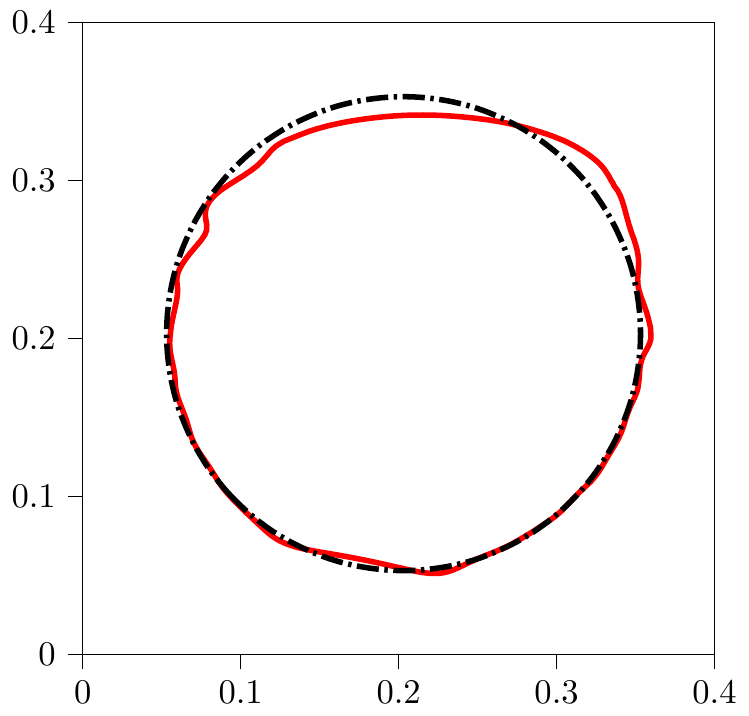}
		\caption*{Resolution : $32^2, p=4$}
	\end{subfigure}

\begin{subfigure}[b]{0.45\linewidth}
\includegraphics{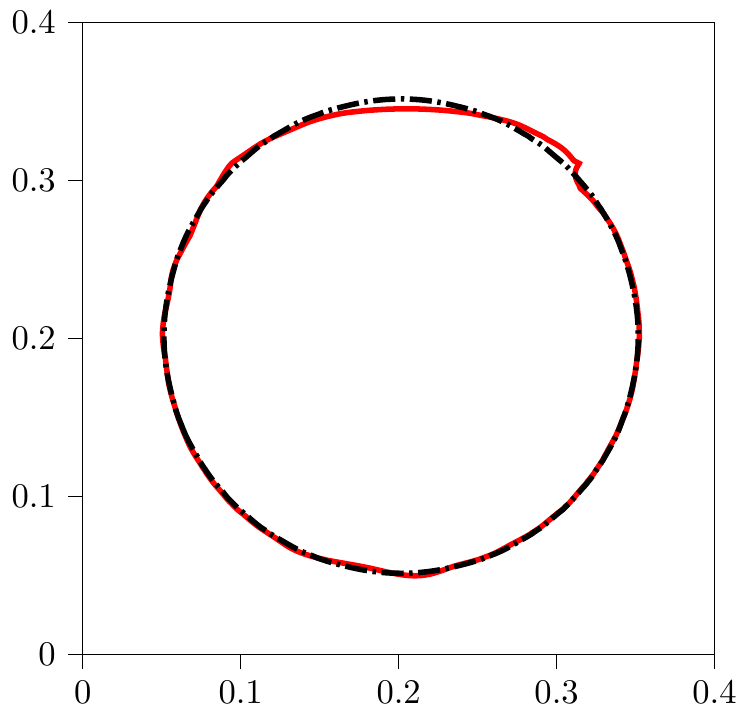}
		\caption*{Resolution : $64^2, p=2$}
\end{subfigure}
\begin{subfigure}[b]{0.45\linewidth}
\includegraphics{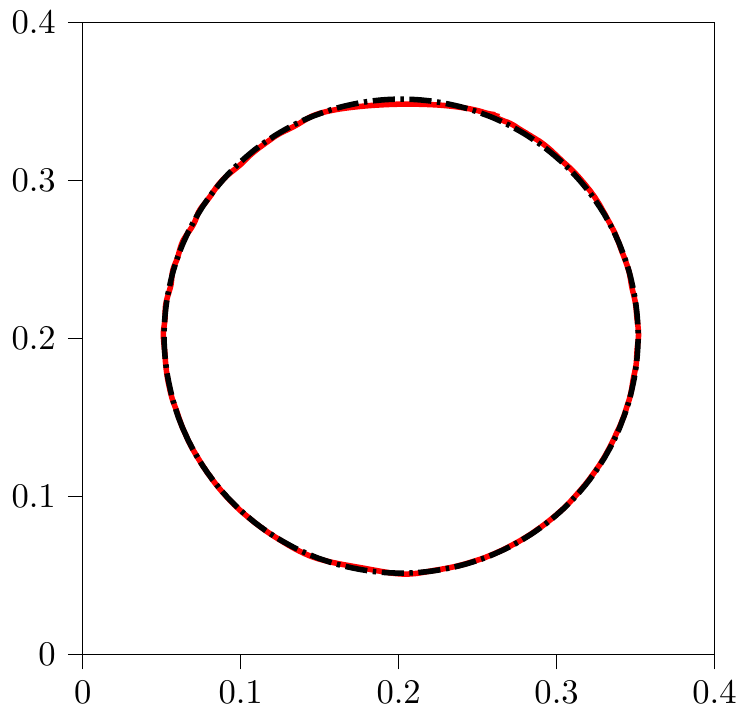}
		\caption*{Resolution : $64^2, p=4$}
\end{subfigure}
\caption{Final (red) and initial (black) $\phi = 0.5$ contour for the Rider-Kothe vortex test using different meshes and polynomial orders}
\label{vortexImprovement}
\end{figure}

Fig. \ref{vortexImprovement} illustrates the improvement of the accuracy with increased mesh resolution and basis orders up to $64^2, \, p=4$ as no further improvement can be discerned visually with finer meshes or higher orders. 

Table \ref{vortexComparison} compares the $E_{L1}$ error obtained from the proposed scheme to some of the best and most recent interface capturing studies. Using $p=2$ is adequate to achieve comparable or superior accuracy when compared to other methods. Consistent improvement in accuracy can be seen when the polynomial order is increased.

\begin{figure}
	\centering
		\begin{subfigure}[b]{0.4\textwidth}
		\includegraphics[width=\textwidth]{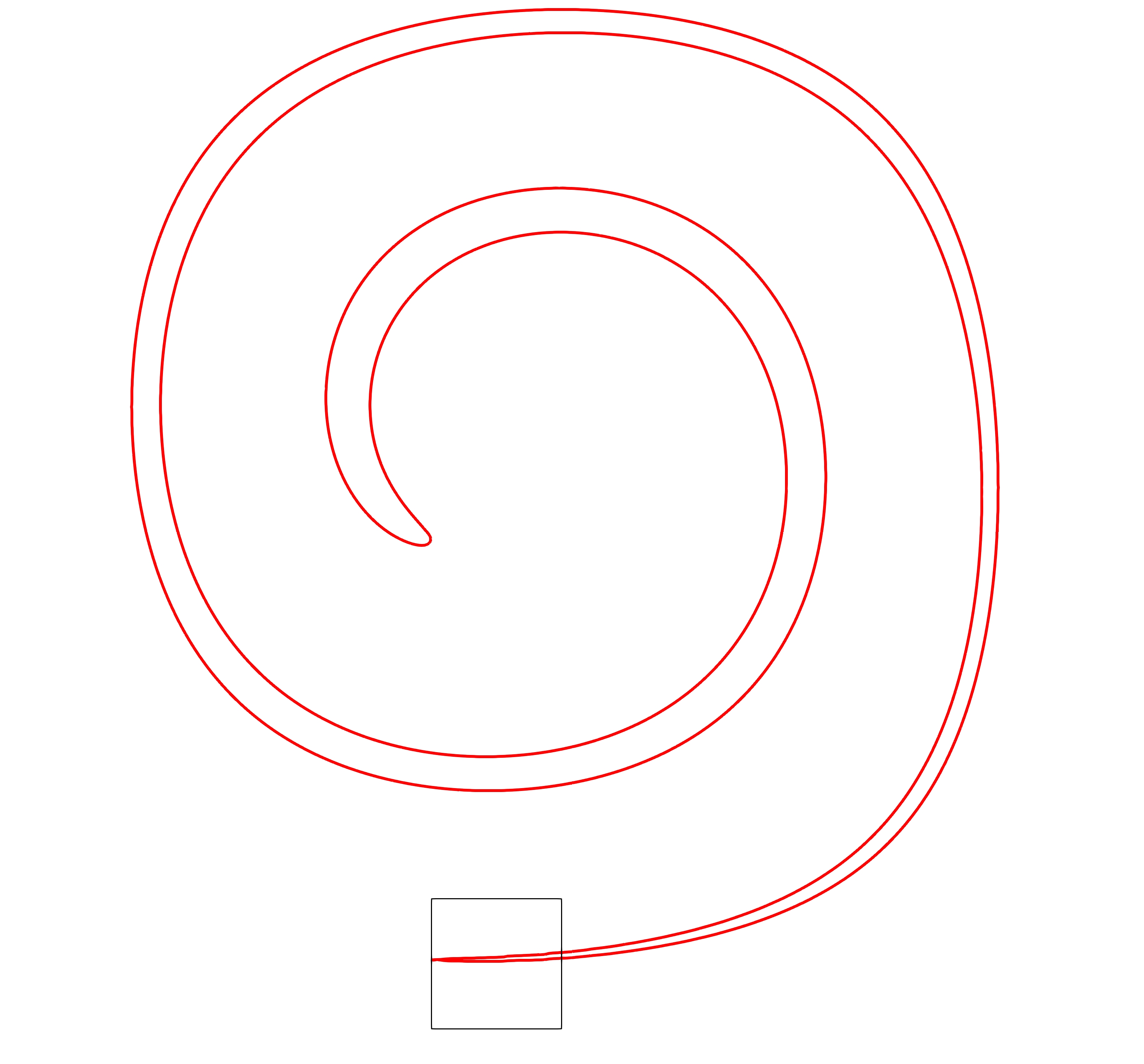}
		\caption*{}
	\end{subfigure}
		\begin{subfigure}[b]{0.45\textwidth}
	\includegraphics[width=\textwidth]{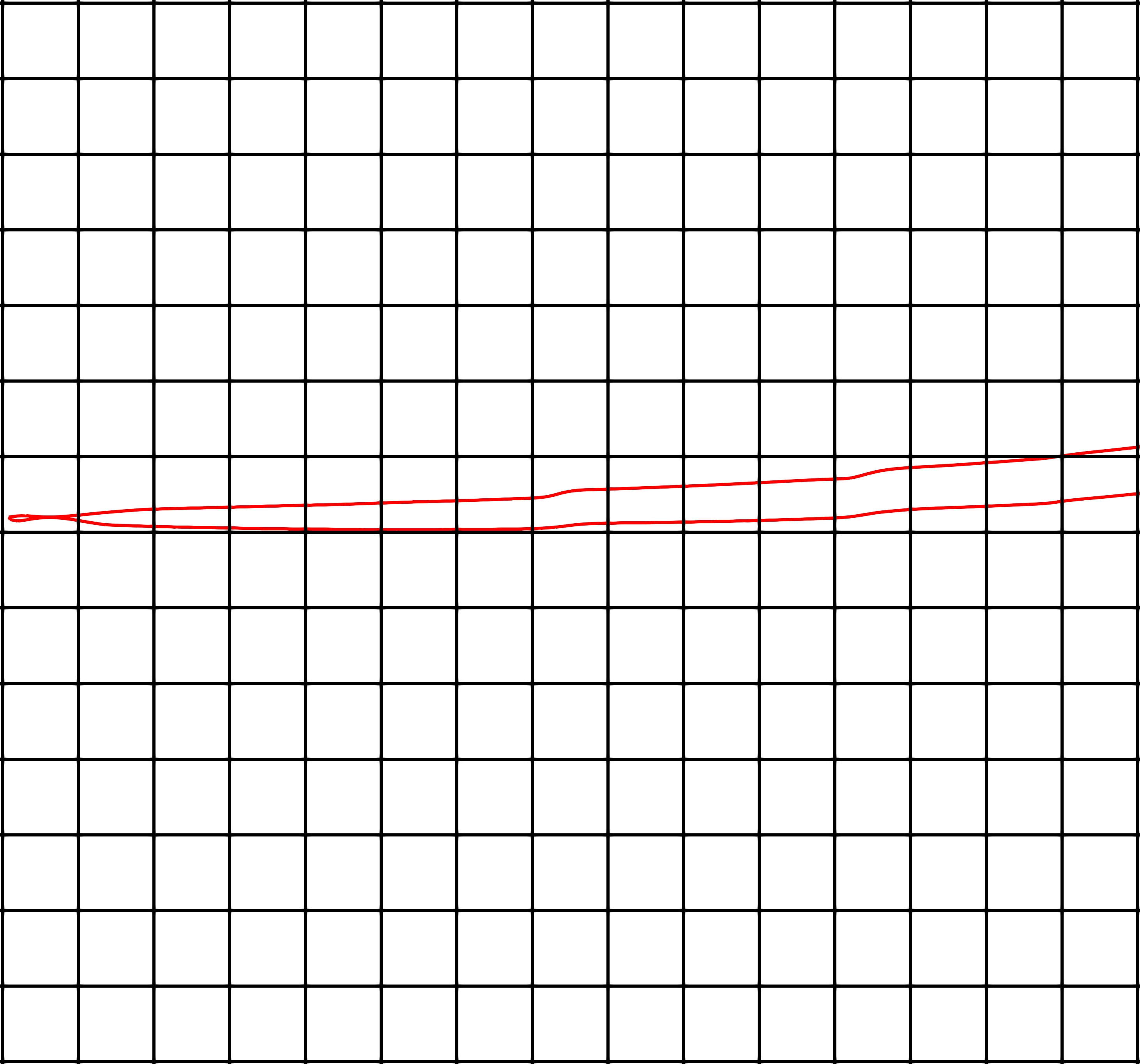}
	\caption*{}
\end{subfigure}
\caption{The Rider-Kothe vortex $ \phi = 0.5 $ contour for the $128^2$, $p=5$ case with an enalrged view of the tail}
\label{tailZoom}
\end{figure}

In \cref{tailZoom} the ability of the FR-PCPF method to resolve sub-grid interface features is demonstrated by providing an enlarged view of the tip of the vortex at  $t=T/2$. Multiple interfaces within an element are resolved without introducing artifacts into the solution. 

\begin{table}
\begin{tabularx}{1.0\textwidth}{L{0.55} C{1.075}C{1.075}C{1.075} C{0.05} C{1.075} C{1.075} C{1.075}}
	\toprule
    \multirow{1}{*}{} &
\multicolumn{3}{c}{Preconditioned CPF} & &
\multicolumn{3}{c}{Unpreconditioned CPF} \\ \cmidrule{2-4}   \cmidrule{6-8} 
\textbf{Mesh}& {$p=3$} & {$p=4$} & {$p=5$} & & {$p=3$} & {$p=4$} & {$p=5$} \\
\midrule \midrule
	
	$b-32^2 $	  &    $7.98 \times 10^{-3}$    &     $4.42 \times 10^{-3}$    & $3.69 \times 10^{-3}$   & &   $1.16\times 10^{-2}$  &   $8.14\times 10^{-3}$  &   $7.91 \times 10^{-3}$ \\
	
	$b-64^2 $	  &    $1.61 \times 10^{-3}$    &     $1.1\times 10^{-3}$    & $7.45 \times 10^{-4}$   & &   $4.11 \times 10^{-4}$   &   $2.43 \times 10^{-3}$  &   $2.16 \times 10^{-3}$ \\
	
	$b-128^2 $	  &    $6.19 \times 10^{-4}$    &     $3.56 \times 10^{-4}$    & $2.41 \times 10^{-4}$   & &   $2.11 \times 10^{-3}$  &   $1.75 \times 10^{-3}$  &   $2.24 \times 10^{-4} $ \\ \\
	
		$c-32^2 $	  &    $6.11\times 10^{-3}$    &     $5.51\times 10^{-3}$    & $3.97 \times 10^{-3}$  & &   $1.24 \times 10^{-2}$  &   $1.07 \times 10^{-2}$  &   $8.18 \times 10^{-3}$ \\
	
	$c-64^2 $	  &    $2.14 \times 10^{-3}$    &     $1.25 \times 10^{-3}$    & $7.64\times 10^{-4}$   & &  $6.65 \times 10^{-3}$   &   $3.8 \times 10^{-3}$  &   $2.81 \times 10^{-3}$ \\
	
	$c-128^2 $	  &    $6.67 \times 10^{-4}$    &     $3.79 \times 10^{-4}$    & $2.6 \times 10^{-4}$   & &  $1.77 \times 10^{-3}$  &   $1.77 \times 10^{-3}$  &   $6.07 \times 10^{-4} $ \\
\bottomrule
\end{tabularx}
	\caption{$E_{L1}$ error for the Rider-Koth vortex on different meshes, resolutions and polynomial orders using the preconditioned and unpreconditioned conservative phase field  equation}
\label{rvortexTable}
\end{table}

\begin{table}
	\begin{tabularx}{1\textwidth}{L{1.5} L{0.833} L{0.833} L{0.833}}
		\toprule
		\textbf{Method}	  &    $32^2$    &     $64^2$    & $128^2$   \\ \midrule \midrule
		Rider-Kothe/Puckett \cite{rider1998reconstructing} &$4.78 \times 10^{-2}$ &$6.96 \times 10^{-3}$&$ 1.44 \times 10^{-3}$ \\
		EMFPA/Puckett \cite{sussman2000coupled} &$3.77 \times 10^{-2}$ &$6.58 \times 10^{-3}$&$ 1.07 \times 10^{-3}$ \\
		THINC/QQ \cite{xie2017toward} &$6.70 \times 10^{-2}$ &$1.52 \times 10^{-3}$&$ 3.06 \times 10^{-3}$ \\
		Markers-VOF\cite{lopez2005improved} &$7.41 \times 10^{-3}$ &$2.12 \times 10^{-3}$&$ 4.27 \times 10^{-4}$ \\
		
		ISLSVOF \cite{Lyras2020} &$4.19\times 10^{-2}$ &$1.43 \times 10^{-3}$&$ 8.36 \times 10^{-4}$ \\
		THINC/LS (P2) \cite{Qian2018} &$1.00\times 10^{-1}$ &$1.22 \times 10^{-2}$&$ 1.20 \times 10^{-3}$ \\
		THINC/LS (P4) \cite{Qian2018} &$2.85 \times 10^{-2}$ &$3.39 \times 10^{-3}$&$ 6.79 \times 10^{-4}$ \\ \\
		
		FR-PCPF($p=2$)  &$7.96\times 10^{-3}$ &$2.51 \times 10^{-3}$&$ 8.10 \times 10^{-4}$ \\
		FR-PCPF($p=3$)  &$7.91\times 10^{-3}$ &$1.29 \times 10^{-3}$&$ 6.72 \times 10^{-4}$ \\
		FR-PCPF($p=4$)  &$5.50\times 10^{-3}$ &$1.05 \times 10^{-3}$&$ 3.75 \times 10^{-4}$ \\
		FR-PCPF($p=5$)  &$3.22\times 10^{-3}$ &$7.46 \times 10^{-4}$&$ 2.64 \times 10^{-4}$ \\

		\bottomrule
	\end{tabularx}
	\caption{$E_{L1}$ error for the Rider-Koth vortex test on quadrilateral mesh }
	\label{vortexComparison}
\end{table}

\begin{figure}[htbp!]
	\centering
\includegraphics{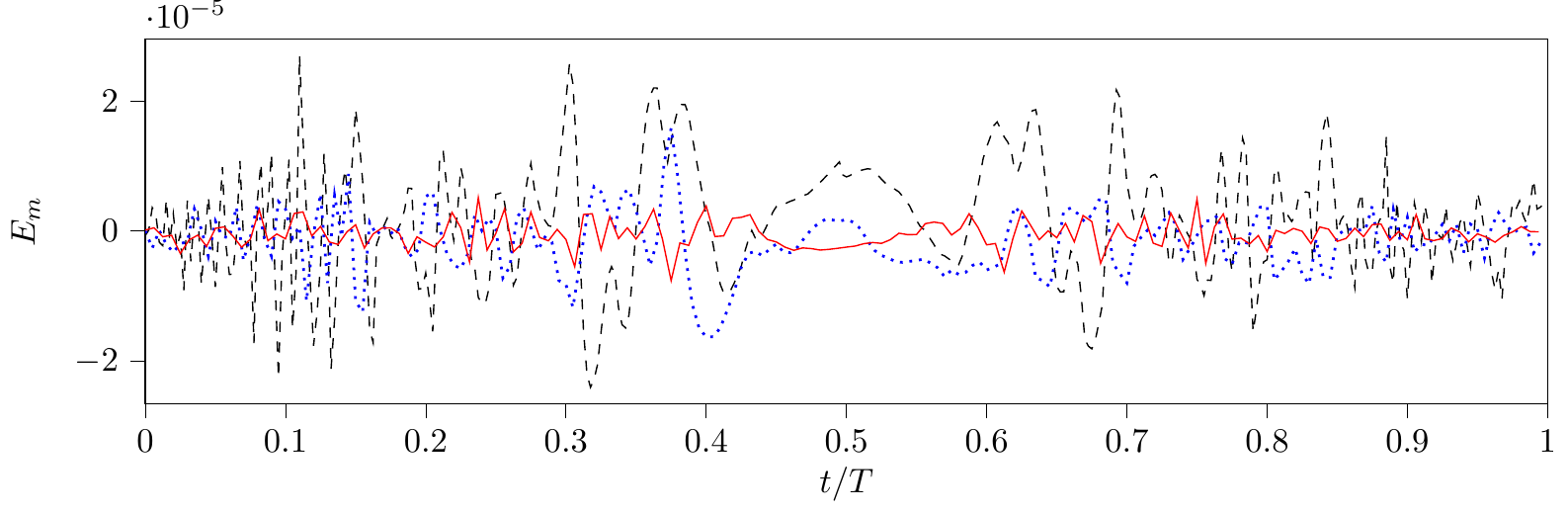}
	\caption{Mass error on a quad $64^2$ mesh for $p=3$(black), $p=4$(blue), $p=5$(red)}
	\label{sheaeringMassLossOrders} 
\end{figure}

\begin{figure}[htbp!]
	\centering
\includegraphics{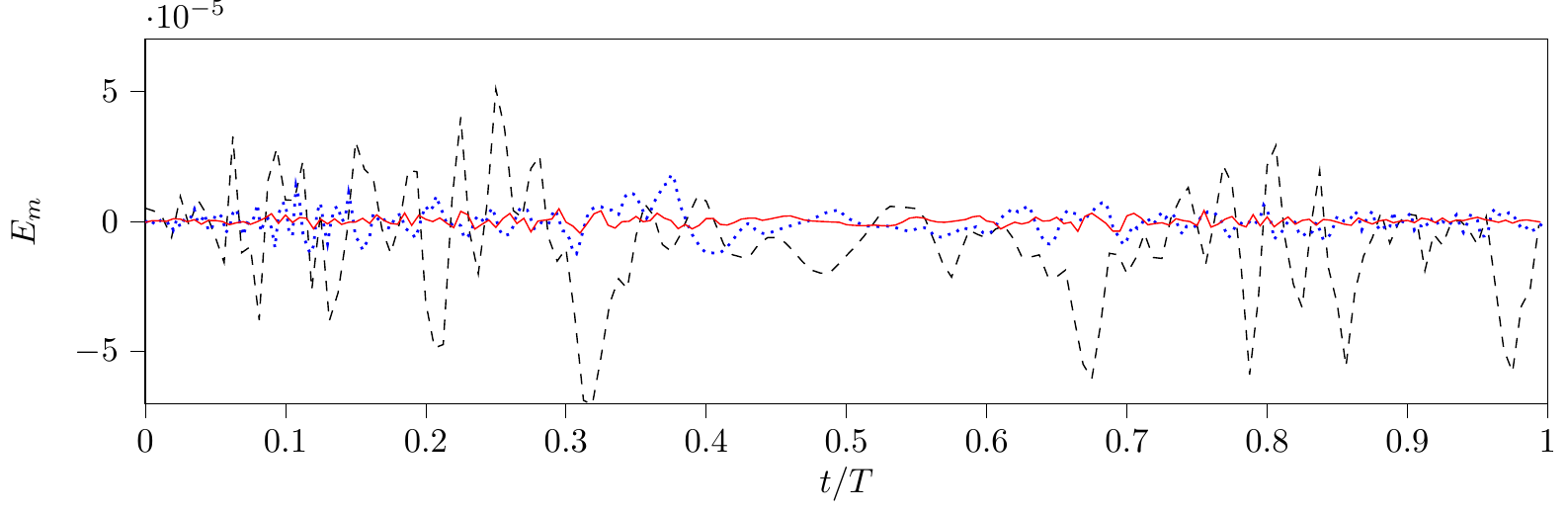}
	\caption{Mass error on $32^2$(black), $64^2$(blue), $128^2$(red) quad meshes using $p=4$}
	\label{sheaeringMassLossMeshes}
\end{figure}

The time history of global mass change ($ E_m $) in the domain is provided in \cref{sheaeringMassLossMeshes,sheaeringMassLossMeshes}. The figures clearly demonstrates improvement in global mass conservation with increased effective resolution ($h/p$).
\subsection{3D Vortex Test}

In order to verify the ability of the proposed method to capture distorted interfaces in 3D, the deformation of a sphere in a 3D velocity field is simulated \cite{leveque1996high}. A sphere with radius $r = 0.15$ in a unit domain centered at $(0.35, 0.35, 0.35)$ is deformed with the following velocity field

\begin{equation}
\begin{split}
u(x,y,t) = 2\sin^2(\pi x) \sin(2\pi y) \sin(2\pi z) \cos (\frac{\pi t}{T}) \\
v(x,y,t) = -\sin(2\pi x) \sin^2(\pi y) \sin(2\pi z) \cos (\frac{\pi t}{T})\\
w(x,y,t) = -\sin(2\pi x) \sin(2\pi y) \sin^2(\pi z) \cos (\frac{\pi t}{T}).
\end{split}
\end{equation}

For all tests in this section, the period $T =3$. Similarly to the Rider-Koth vortex, the sphere is deformed until $T/2$ and then the velocity field reverses to restore the sphere to its initial condition at $t=T$. This test was done on hexahedral meshes with $32^3$, $64^3$ and $128^3$ elements using $p2 - p5$. The large number of solution points needed to carry out these tests (up to $0.45$ billion points for the $128^3$ mesh with $p=5$ case) necessitated using as many as 40 Nvidia P100 GPUs to accommodate the large computational resources needed.

\begin{figure}[h]
	\centering
	\begin{subfigure}[b]{0.2\textwidth}
		\includegraphics[width=\textwidth]{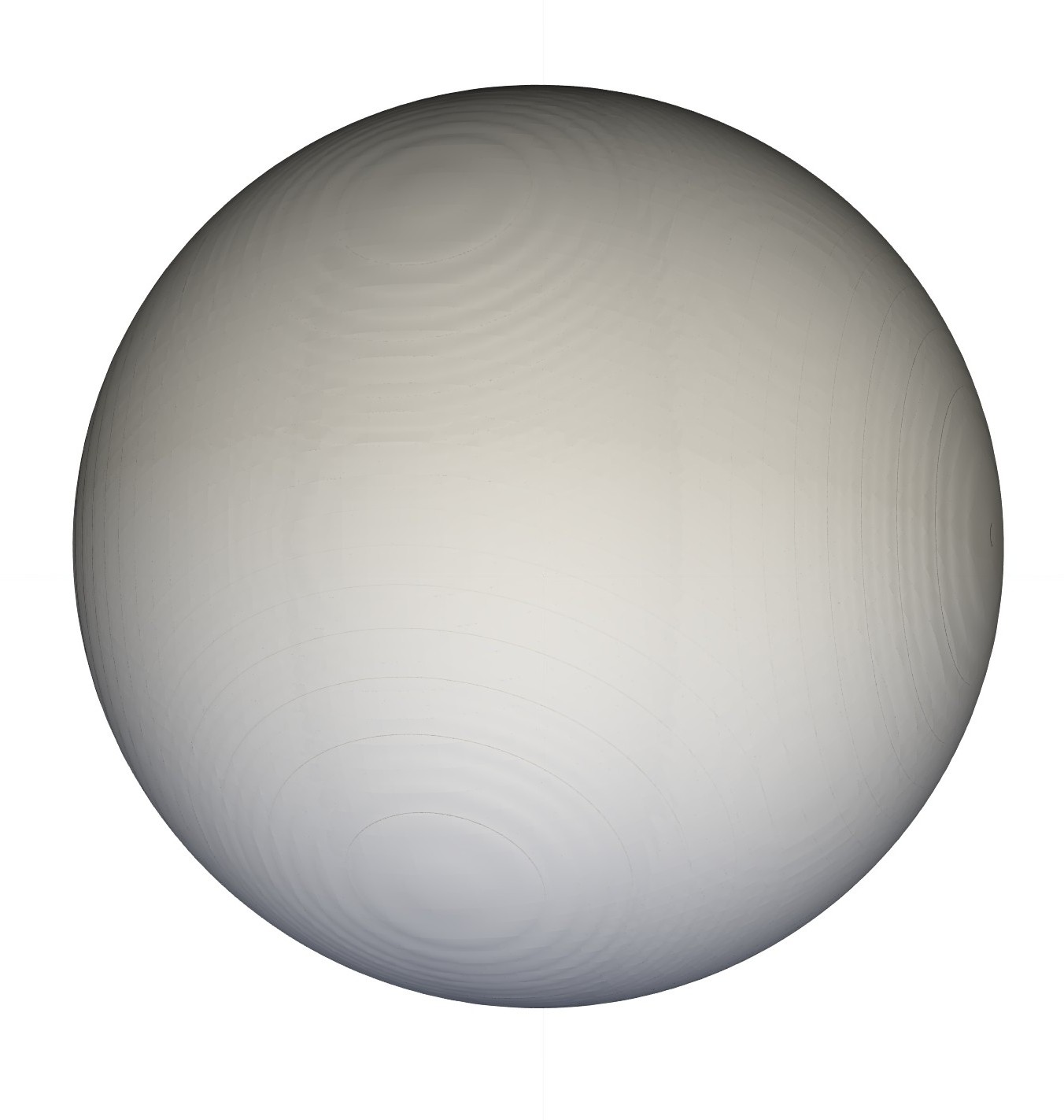}
		\caption*{t=0.0}
	\end{subfigure}

	\begin{subfigure}[b]{0.19\textwidth}
	\includegraphics[width=\textwidth]{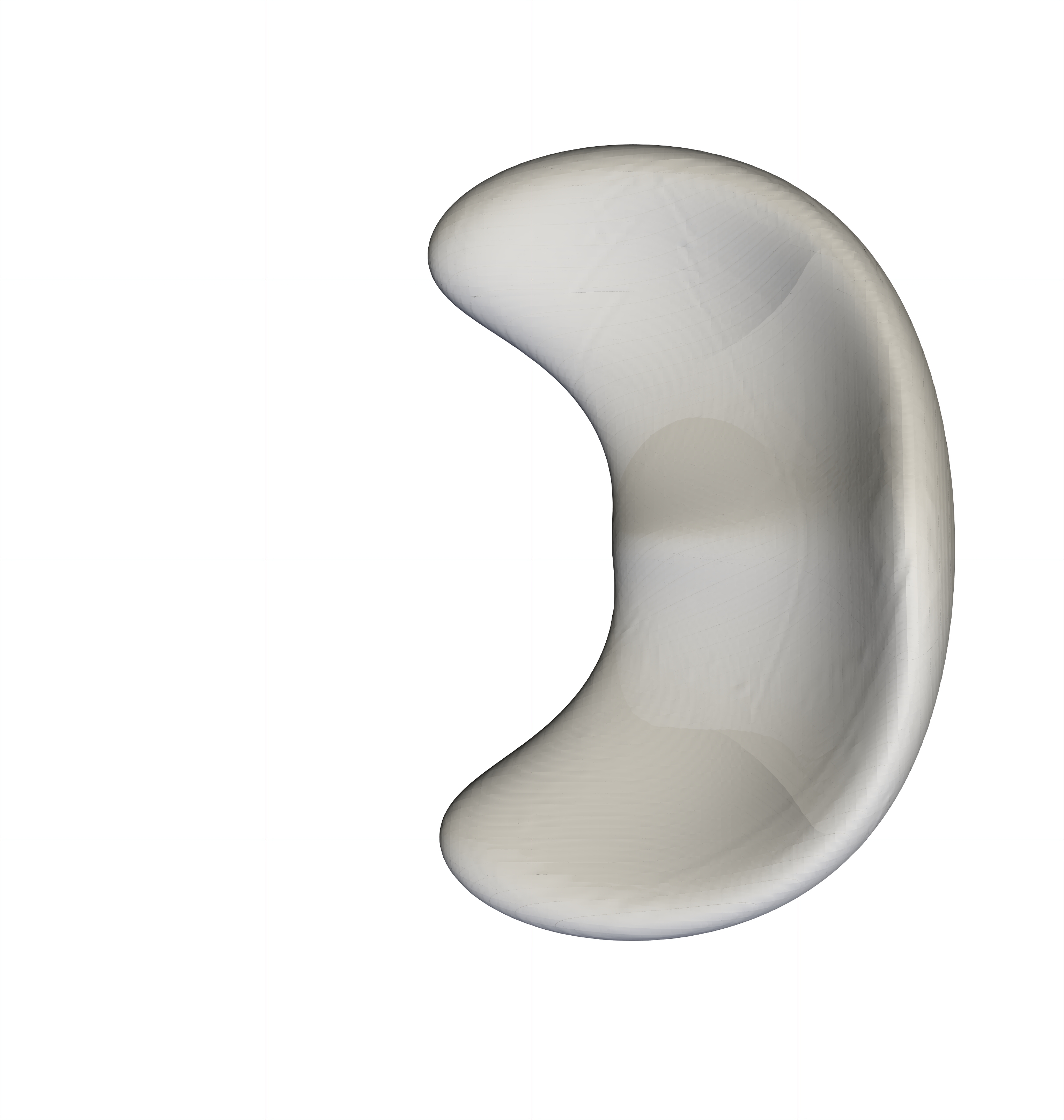}
		\caption*{t=0.5}
\end{subfigure}
	\begin{subfigure}[b]{0.19\textwidth}
	\includegraphics[width=\textwidth]{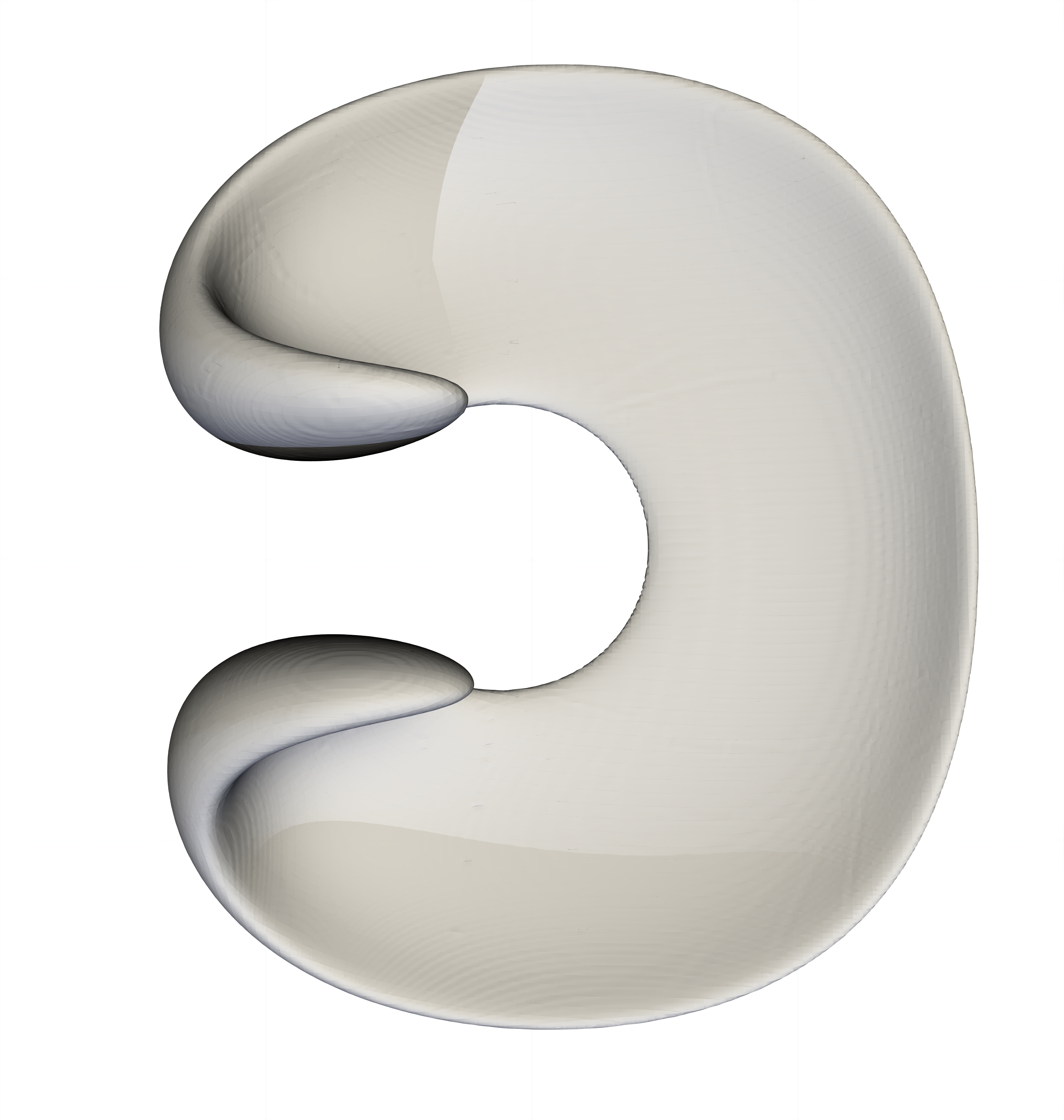}
		\caption*{t=1.0}
\end{subfigure}
	\begin{subfigure}[b]{0.19\textwidth}
	\includegraphics[width=\textwidth]{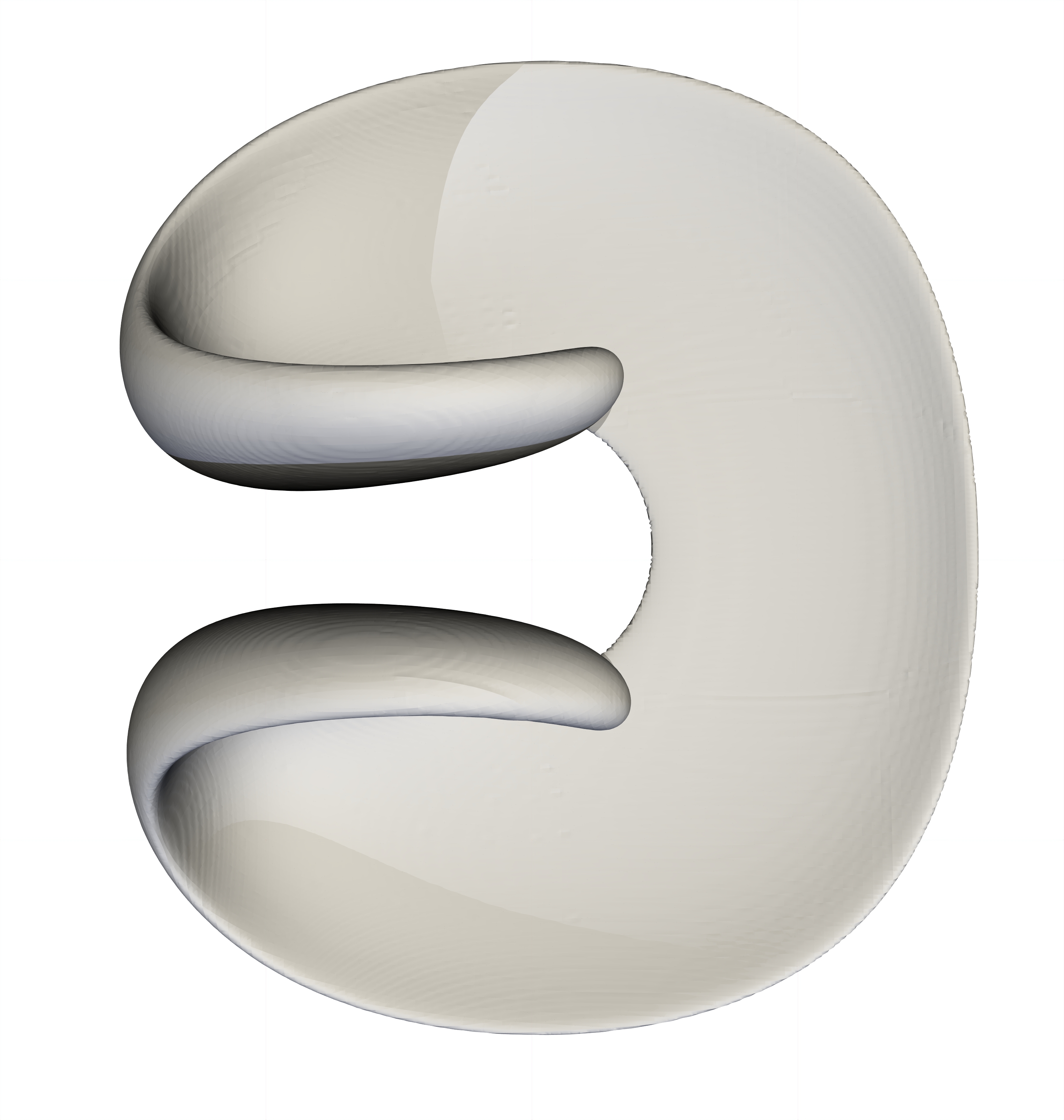}
		\caption*{t=1.5}
\end{subfigure}
	\begin{subfigure}[b]{0.19\textwidth}
	\includegraphics[width=\textwidth]{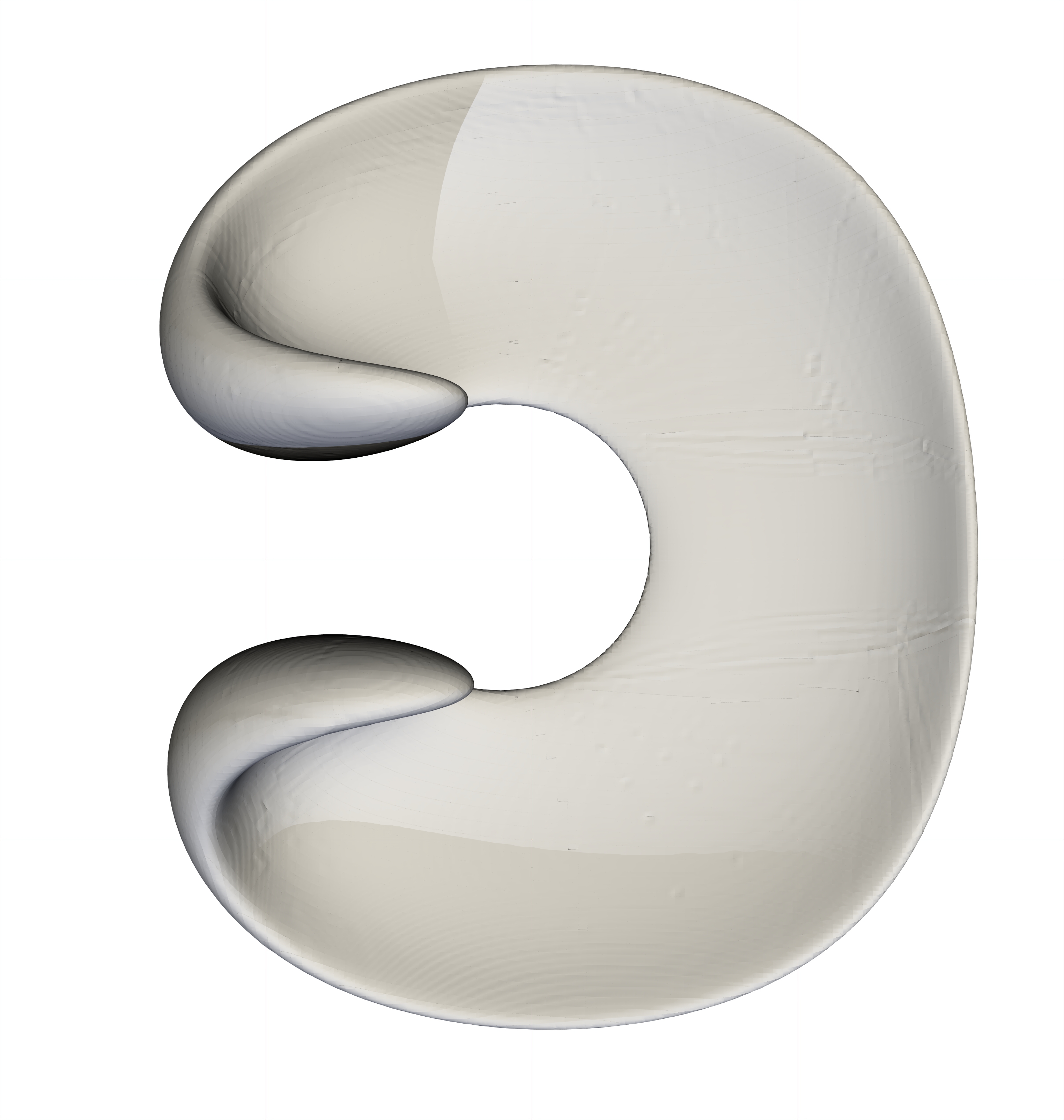}
		\caption*{t=2.0}
\end{subfigure}
	\begin{subfigure}[b]{0.19\textwidth}
	\includegraphics[width=\textwidth]{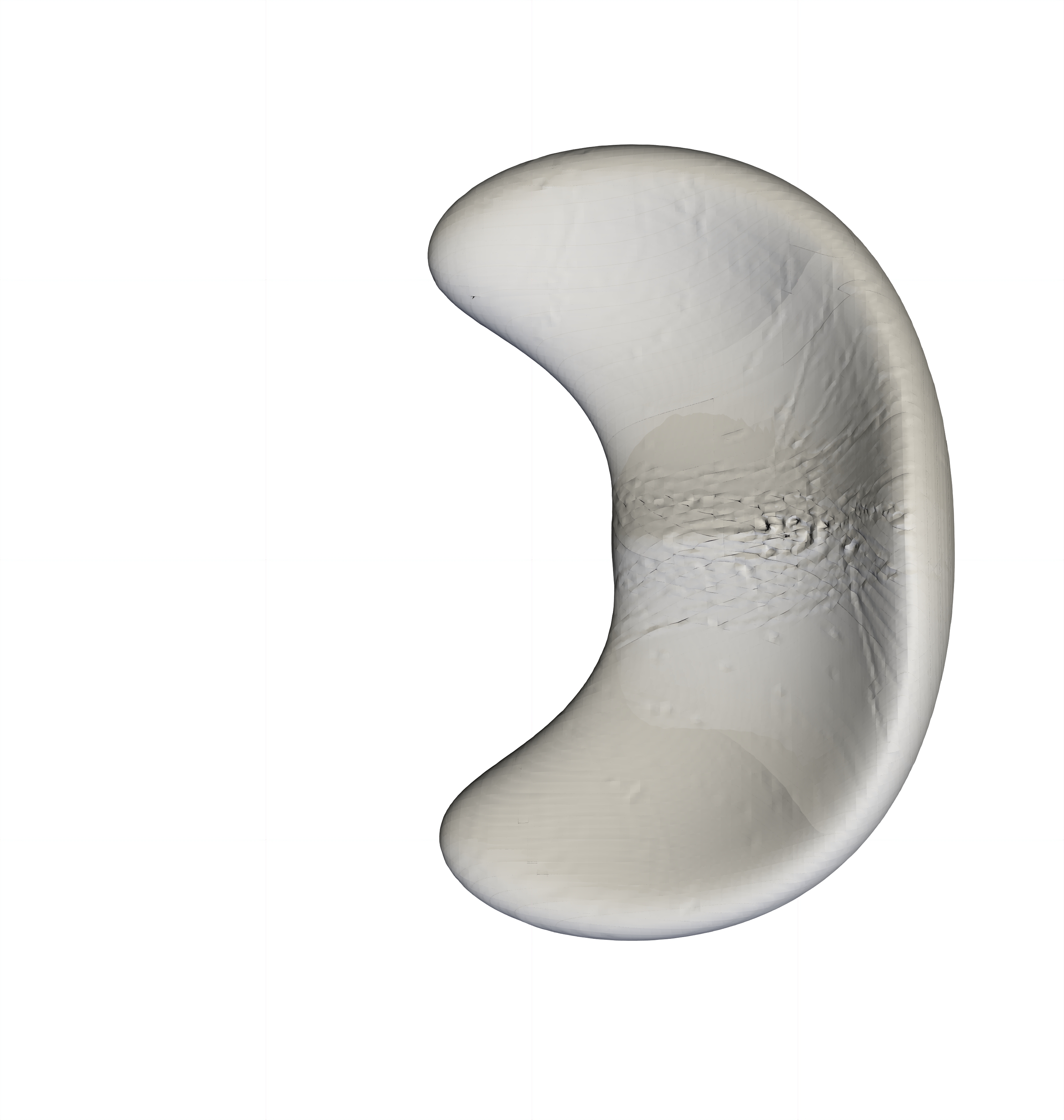}
		\caption*{t=2.5}
\end{subfigure}

	\begin{subfigure}[b]{0.2\textwidth}
	\includegraphics[width=\textwidth]{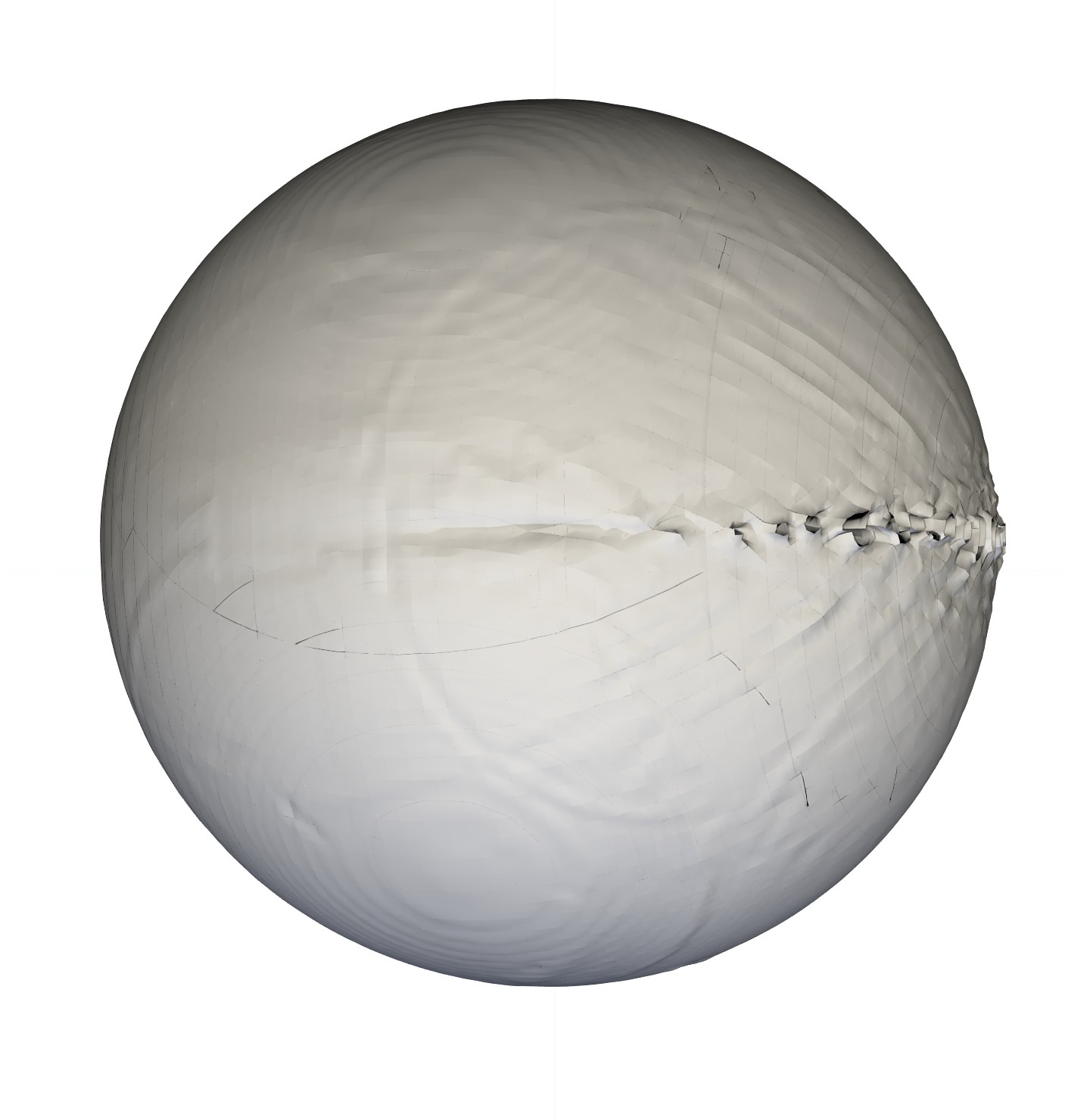}
		\caption*{t=3}
\end{subfigure}
	\caption{Time evolution of the $\phi = 0.5$ contour of the 3D vortex problem}
	\label{threeDVortexHistory}
\end{figure}

The evolution of the sphere for the $128^3$ mesh case and $p=4$ is depicted in \cref{threeDVortexHistory}. The figure shows some minor artifacts at the interface while remaining quantitatively accurate as highlighted by the values of $E_{L1}$ error comparison outlined in  \Cref{threeDimvortexComparison}. The table further shows that the FR-PCPF method is able to produce more accurate results even in comparison to complex methods that rely on geometric reconstruction or employ high-order schemes that are also able resolve sub-cell interface structures. 
\begin{table}[h]
	\begin{tabularx}{1\textwidth}{L{1.5} L{0.833} L{0.833} L{0.833}}
		\toprule
		\textbf{Method}	  &    $32^2$    &     $64^2$    & $128^2$   \\ \midrule \midrule
		LVIRA \cite{jofre20143} &$6.92 \times 10^{-3}$ &$2.43 \times 10^{-3}$&$ 6.37 \times 10^{-4}$ \\		
		THINC/QQ \cite{xie2017toward} &$7.96 \times 10^{-3}$ &$2.89 \times 10^{-3}$&$ 9.05 \times 10^{-4}$ \\
		DS-CLSMOF \cite{jemison2013coupled} &$4.81 \times 10^{-3}$ &$1.99 \times 10^{-3}$&$ -$ \\
		
		ISLSVOF \cite{Lyras2020} &$8.89\times 10^{-3}$ &$2.96 \times 10^{-3}$&$ 8.06 \times 10^{-4}$ \\
		THINC/LS (P2) \cite{Qian2018} &$6.81\times 10^{-3}$ &$2.07 \times 10^{-3}$&$ 4.51 \times 10^{-4}$ \\
		THINC/LS (P4) \cite{Qian2018} &$5.54 \times 10^{-3}$ &$1.57 \times 10^{-3}$&$ 3.79 \times 10^{-4}$ \\ \\
		
		FR-PCPF($p=2$)  &$6.56\times 10^{-3}$ &$2.56 \times 10^{-3}$&$ 6.61 \times 10^{-4}$ \\
		FR-PCPF($p=3$)  &$3.05\times 10^{-3}$ &$8.89 \times 10^{-4}$&$ 4.52 \times 10^{-4}$ \\
		FR-PCPF($p=4$)  &$2.28\times 10^{-3}$ &$6.49 \times 10^{-4}$&$ 2.05 \times 10^{-4}$ \\
		FR-PCPF($p=5$)  &$1.67\times 10^{-3}$ &$4.85 \times 10^{-4}$&$ 1.60 \times 10^{-4}$ \\
		
		\bottomrule
	\end{tabularx}
	\caption{$E_{L1}$ error for the 3D vortex test on hexahedral meshes }
	\label{threeDimvortexComparison}
\end{table}

\section{Conclusions}

To pave the way to high-order multi-phase simulations using the Flux Reconstruction method, we develop a novel interface capturing approach. The merits and novelties of the proposed method can be summarized as follows:

\begin{itemize}
	\item To enable capturing sharp oscillation-free interfaces using a high-order approach, we develop a localized conservative phase field method.
	\item The accuracy of the phase field method is significantly improved by introducing a level-set based preconditioning step that increases the accuracy of interface normals, and prevents the appearance of smearing and fragmentation artifacts.	
	\item The favorable numerical properties of the FR-PCPF approach, such as the high-order polynomial representation of all field variables, makes it possible to capture sub-grid interface features with a high level of accuracy.
\end{itemize}

Numerical validation test cases have shown that using second order polynomial basis is sufficient to produce results that are comparable to some of the most accurate existing methods. Increasing the polynomial order to higher than 3 consistently improves numerical accuracy. The FR-PCPF method is shown to be able to maintain good mass conservation and resolve challenging interface features even with long time integration. Increasing effective resolution $ (h/p) $ was shown to produce consistent improvements in global mass conservation, accuracy and the ability to resolve complicated interface features. 

\clearpage
\bibliography{sources}
\bibliographystyle{ieeetran}

\end{document}